\documentclass{article}
\usepackage{flushend} 
\usepackage{balance}

\usepackage{array,multirow,graphicx}
\usepackage{microtype}
\usepackage{graphicx}
\usepackage{subfigure}
\usepackage{booktabs} 

\usepackage{hyperref}

\usepackage[accepted]{sty}
\usepackage{graphicx}
\usepackage{pythonhighlight}

\usepackage{times}
\usepackage{helvet}
\usepackage{courier}
\usepackage{subfigure} 
\usepackage{cancel}
\RequirePackage{amsthm,amsmath}
\usepackage{graphicx, amssymb, mathrsfs,amsfonts,url, bm}

\newcommand{\real}{\mathbb{R}}

\newcommand{\gap}{\,\,\,\,\,\,\,\,}

\newcommand{\bSigma}{\boldsymbol\Sigma}

\newcommand{\bA}{\bm{A}}

\newcommand{\bM}{\bm{M}}

\newcommand{\bp}{\bm{p}}

\newcommand{\br}{\bm{r}}

\newcommand{\bw}{\bm{w}}

\newcommand{\bx}{\bm{x}}
\newcommand{\bX}{\bm{X}}
\newcommand{\by}{\bm{y}}
\newcommand{\bY}{\bm{Y}}
\newcommand{\bz}{\bm{z}}

\usepackage[font=small,skip=2pt]{caption}
 
\newcommand{\titlethis}{Autoencoding Conditional GAN for Portfolio Allocation Diversification}

\icmltitlerunning{\titlethis}

\begin{document}

\twocolumn[
\icmltitle{\titlethis}

\begin{icmlauthorlist}
	\icmlauthor{Jun Lu}{te}
	\icmlauthorsingle{\gap \gap Shao Yi}
\end{icmlauthorlist}

\icmlaffiliation{te}{Correspondence to: Jun Lu $<$jun.lu.locky@gmail.com$>$. Copyright 2022 by the author(s)/owner(s). June 18th, 2022}



\vskip 0.3in
]



\printAffiliationsAndNotice{}  

\begin{abstract}
Over the decades, the Markowitz framework has been used extensively in portfolio analysis though it puts too much emphasis on the analysis of the market uncertainty rather than on the trend prediction.
While generative adversarial network (GAN) and conditional GAN (CGAN) have been explored to generate financial time series and extract features that can help portfolio analysis. The limitation of the CGAN framework stands in putting too much emphasis on generating series rather than keeping features that can help this generator.
In this paper, we introduce an autoencoding CGAN (ACGAN) based on deep generative models that learns the internal trend of historical data while modeling market uncertainty and future trends. 
We evaluate the model on several real-world datasets from both the US and Europe markets, and show that the proposed ACGAN model leads to better portfolio allocation and generates series that are closer to true data compared to the existing Markowitz and CGAN approaches.

%

\end{abstract}
\section{Introduction}

Financial portfolio management is largely based on linear models and the Markowitz framework \citep{markowitz1968portfolio, markowitz1976markowitz} though the underlying data and information in today's market has increased countless times over that of many years ago. The fundamental idea behind the Markowitz framework is to create portfolio diversification while reducing specific risks and assessing the risk-return trade-offs for each asset.
The Markowitz framework, on the other hand, has been criticized for making ideal assumptions about the financial system and data: the expected mean returns, and the covariance matrix of the return series are estimated from the past observations and assumed constant in the future. However, this is such a strong assumption that the market will find it impossible to meet this requirement in practice.

The classic portfolio assessment approach calculates portfolio risk indicators based on asset price series in the past period, such as variance, value at risk, and expected loss, as the evaluation results of cross-section risk.
However, the traditional method has two obvious drawbacks. 
First, because the capital market is changing rapidly, historical data usually cannot be used to indicate the situation in the future; when a trusted long-term forecast is offered in a high efficient market, this prediction is absorbed by traders in the short term and has a direct impact on current price, while future price variations are unpredictable again \citep{timmermann2004efficient}. 
Secondly, the risk measurement indicators estimated by traditional methods usually only contain the linear components in the historical series, leaving out the nonlinear information contained therein, resulting in the deviation between the evaluation results and the real situation \citep{tsay2005analysis}.

On the other hand, the financial market is one of the most heavily impacted industries by AI advancements. Machine learning has been used in a variety of applications, including series generation, forecasting, customer service, risk management, and portfolio management \citep{huang2005forecasting, kara2011predicting, takahashi2019modeling}.
Specifically, generative adversarial networks (GANs) are a sort of neural network architectures that have shown promise in image generation and are now being used to produce time series and other financial data \citep{goodfellow2014generative, esteban2017real, eckerli2021generative}. 
There are several techniques to model financial time series data, including models of the ARCH and GARCH family, which use classical statistics to model the change in variance over time in a time series by characterizing the variance of the current error component as a function of previous errors \citep{engle1982autoregressive, bollerslev1986generalized, lu2022reducing}.
GANs are being used to address some of the paucity of real data, as well as to optimize portfolios and trading methods which achieve better results \citep{takahashi2019modeling, mariani2019pagan}.

However, due to its highly noisy, stochastic, and chaotic nature, market price forecasting is still one of the key issues in the time series literature. 
While previous work have tried to generate financial data based on historical trend, the internal features of the past series are not captured sufficiently so that the generated series is not close to the real market trend \citep{mariani2019pagan}.

In this light, we focus on GANs for better portfolio allocation that can both capture historical trends and generate series based on past data. 
We present a novel framework about portfolio analysis based on conditional GAN (CGAN) that incorporates autoencoder, hence the name \textit{autoencoding CGAN (ACGAN)}, to overcome the issues and challenges encountered in portfolio management tasks. 
Similar to the CGAN model for portfolio analysis \citep{mariani2019pagan}, ACGAN can also directly model the market uncertainty via its complex multidimensional form, which is the primary driver of future price trends, such that the nonlinear interactions between various assets can be embedded effectively. 
We assess the proposed ACGAN method on two separate portfolios representing different markets (the US and the European markets) and industrial segments (e.g., Healthcare, Technology, Industrials, and Basic materials sectors). 
The empirical results show that the proposed approach is capable of realizing the risk-return trade-off and outperforms the classic Markowitz framework and CGAN-based methodology considerably.

\section{Related Work}
As aforementioned, there are several methods delving with portfolio allocation, including the Markowitz framework and the CGAN methodology \citep{markowitz1968portfolio, mariani2019pagan}. 
The Markowitz framework relies on the assumption that the past trend can be applied in the future. While the CGAN methodology partly solves the drawback in the Markowitz framework by simulating future data based on historical trends, it still lacks full ability to capture the information and features behind the past data. The proposed ACGAN model introduces an extra \textit{decoder} that can help to construct the networks that both capture historical features and simulate data closer to real ones.
\subsection{Markowitz Framework}
Portfolio allocation is a kind of investment portfolio where the market portfolio has highest Sharpe ratio (SR) given the composition of assets \citep{markowitz1968portfolio}. For simplicity, we here only consider long only portfolio. Denote $\br$ as the return on assets vector, $\bSigma$ as the asset covariance matrix, $\bw$ as the weight vector of each asset, and $r_f$ as the risk-free interest rate. If we measure portfolio risk by variance (or standard deviation),
then the overall return and risk of the portfolio are:
\begin{equation}
\begin{aligned}
		r_p &= \bw^\top \br,\\
	\sigma_p^2& = \bw^\top\bSigma\bw.
\end{aligned}
\end{equation}
And the Sharpe ratio \citep{sharpe1966mutual} can be obtained by 
\begin{equation}
\text{SR} =\frac{r_p - r_f}{\sigma_p} = \frac{\bw^\top\br - r_f}{\sqrt{\bw^\top\bSigma\bw}}.
\end{equation}
According to the definition of portfolio allocation, the weight of each asset in the market portfolio is the solution to the following optimization problem:
\begin{equation}
\begin{aligned}
&\mathop{\arg\max}_{\bw} \frac{\bw^\top\br - r_f}{\sqrt{\bw^\top\bSigma\bw}};\\	
&\text{s.t. } \sum_{i=1}^{N} w_i=1; \\
& 0\leq w_i\leq 1, \forall \, i\in \{1,2,\ldots,N \},
\end{aligned}
\end{equation}
where $N$ is the number of assets, and $w_i$ is the $i$-th element of the weight vector $\bw$.

\begin{figure}[t!]
\centering
\includegraphics[width=0.3\textwidth]{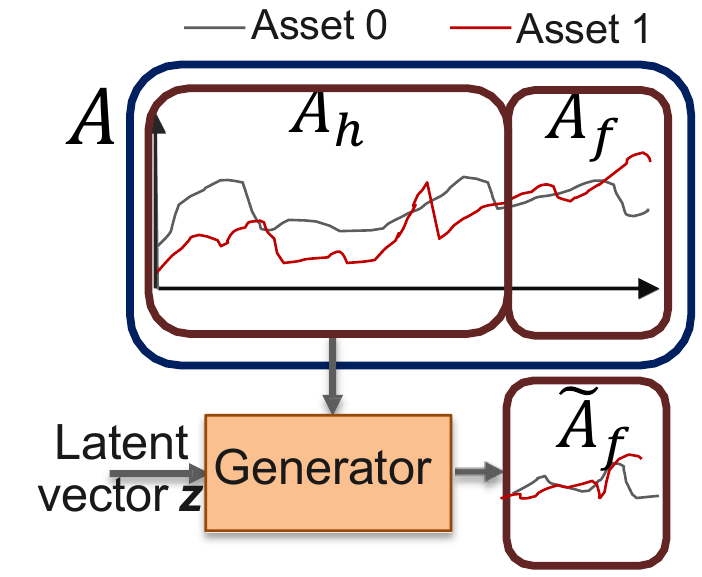}
\caption{A conceptual overview of the CGAN and the proposed ACGAN generators' inputs and outputs.}
\label{fig:overview_of_GAN-series}
\end{figure}

\subsection{Portfolio Analysis with GAN}

We consider the matrix $\bA$ to span the whole analysis length: $w=h+f$. The matrix $\bA$ contains two components, the known historical series $\bA_h$ of length $h$, and the unknown future $\bA_f$ of length $f$.
Given the number of assets $N$, the matrix $\bA$ is of size $N\times w$; $\bA_h$ has shape $N\times h$; and $\bA_f$ is of shape $N\times f$.


Given the known recent historical series $\bA_h\in \real^{N\times h}$ and a prior distribution of a random latent vector $\bz$ (or size $m$: $\bz\in\real^m$), we use a generative deep-neural network $G$ to learn the probability distribution of future price trends $\bA_f$ within the target future horizon $f$.
Figure~\ref{fig:overview_of_GAN-series} depicts a graphical representation of the matrix $\bA$, as well as the generator $G$'s inputs and outputs.
Formally the generative model generates a fake future matrix $\widetilde{\bA}_f$ by 
\begin{equation}
	\widetilde{\bA}_f = G( \bz,\bA_h),
\end{equation}
where $\bz\in \real^m$ is the latent vector sampled from a prior distribution (e.g., from a normal distribution). In practice, the latent vector $\bz$ represents the unforeseeable future occurrences and phenomena that will have an impact on the marketplace. Based on the most recent market conditions, the known historical series $\bA_h$ is used to extract features and condition the probability distribution of the future $\bA_f$.
Given the historical observation $\bA_h$ and following the Wasserstein GAN-GP (WGAN-GP) by \citet{gulrajani2017improved}, the generative $G$ is trained in adversarial mode against a discriminator network $D$ with the goal of minimizing the Wasserstein distance between the real future series $\bA_f$ and the fake series $\widetilde{\bA}_f$. Formally, the process is described by the following optimization problem:
\begin{equation}\label{equation:cgan_losses}
\begin{aligned}
&\mathop{\max}_{D} &\,&\mathbb{E}_{\bx\sim p(\text{data})}
\bigg\{D(\bx) - \mathbb{E}_{\bz\sim p(\bz)}\big[D(G(\bz,\bx_h))\big] \bigg\}
-\\
&& &\,\, \,\lambda_1 \cdot \mathbb{E}_{\overline{\bx}\sim p(\epsilon\, \text{data} + (1-\epsilon) G(\bz))}
\big[ ||\nabla_{\overline{\bx}} D(\overline{\bx}) ||_2 -1 \big]^2;\\
&\mathop{\max}_{G} &\, &\mathbb{E}_{\bx\sim p(\text{data})} \bigg\{\mathbb{E}_{\bz\sim p(\bz)}\big[ D(G(\bz, \bx_h))  \big]\bigg\},\\
\end{aligned}
\end{equation}
where $\bx_h$ contains the historical parts of the data $\bx$ ($\bx_h\in \bx$),  $G(\bz,\bx_h)$ indicates that the generator depends on the (historical) data $\bx_h$, and $\lambda_1$ controls the gradient penalty.
Theoretically, the optimization process finds the surrogate posterior probability distribution $p(\widetilde{\bA}_f | \bA_h)$ that approximates the real posterior probability distribution $p({\bA}_f | \bA_h)$.

\begin{figure}[h]
\centering  
\subfigtopskip=2pt 
\subfigbottomskip=2pt 
\subfigcapskip=-2pt 
\subfigure[Generator]{\includegraphics[width=0.425\textwidth]{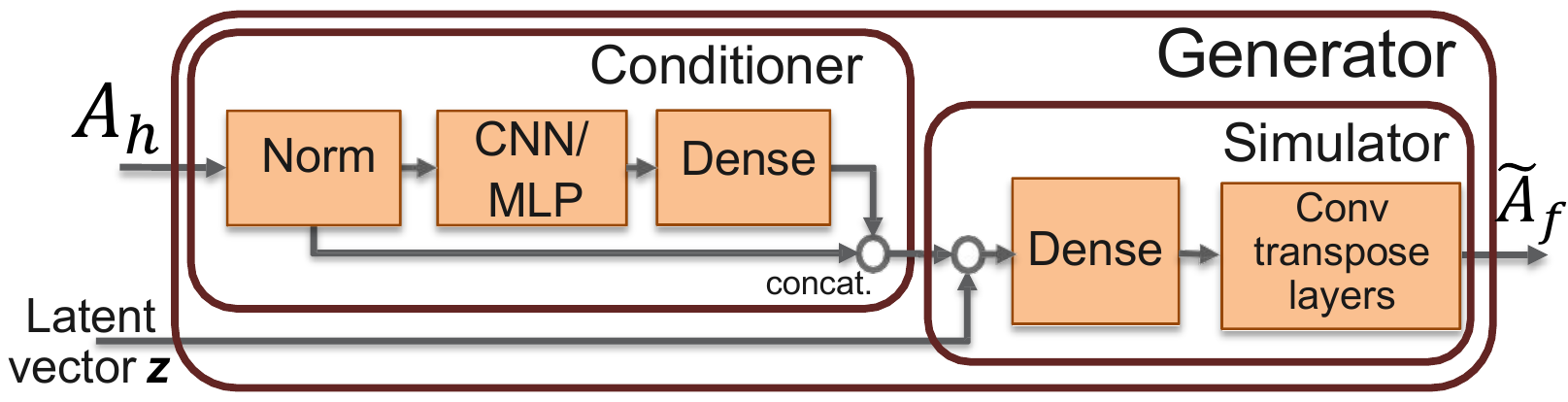} \label{fig:generator_cgan}}
\subfigure[Distriminator]{\includegraphics[width=0.325\textwidth]{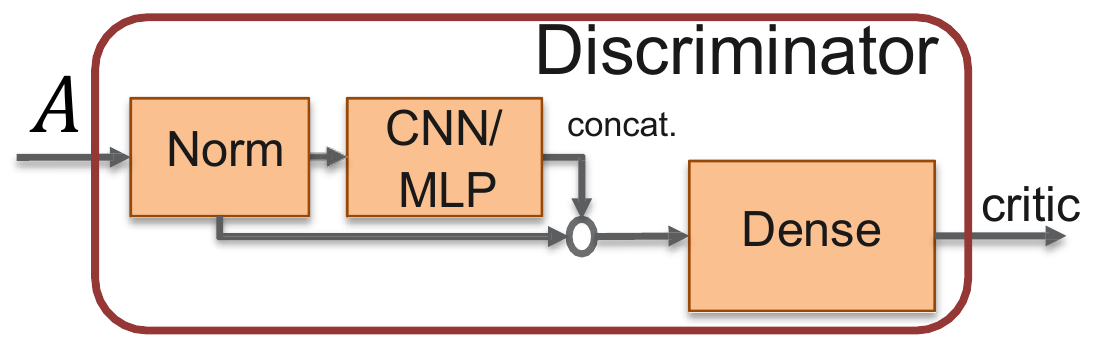} \label{fig:discriminator_cgan}}
\caption{Architectures of the CGAN generative and discriminative models for portfolio analysis.}
\label{fig:cgan_structure}
\end{figure}

The main drawback of the CGAN methodology is in that it puts too much emphasis on the \textit{conditioner} to extract features that can ``deceive" the discriminator (Figure~\ref{fig:generator_cgan}). When the discriminator is perfectly trained, this issue is not a big problem. However, in most cases, especially due to the scarcity of financial data, the discriminator works imperfectly such that the conditioner may lose important information for the historical data.
Whereas in the proposed ACGAN model, we find a balance between the information extraction and generation for cheating the discriminator via an embedded autoencoder providing the capability of keeping the intrinsic information of historical data.

\begin{figure}[t!]
	\centering
	\includegraphics[width=0.425\textwidth]{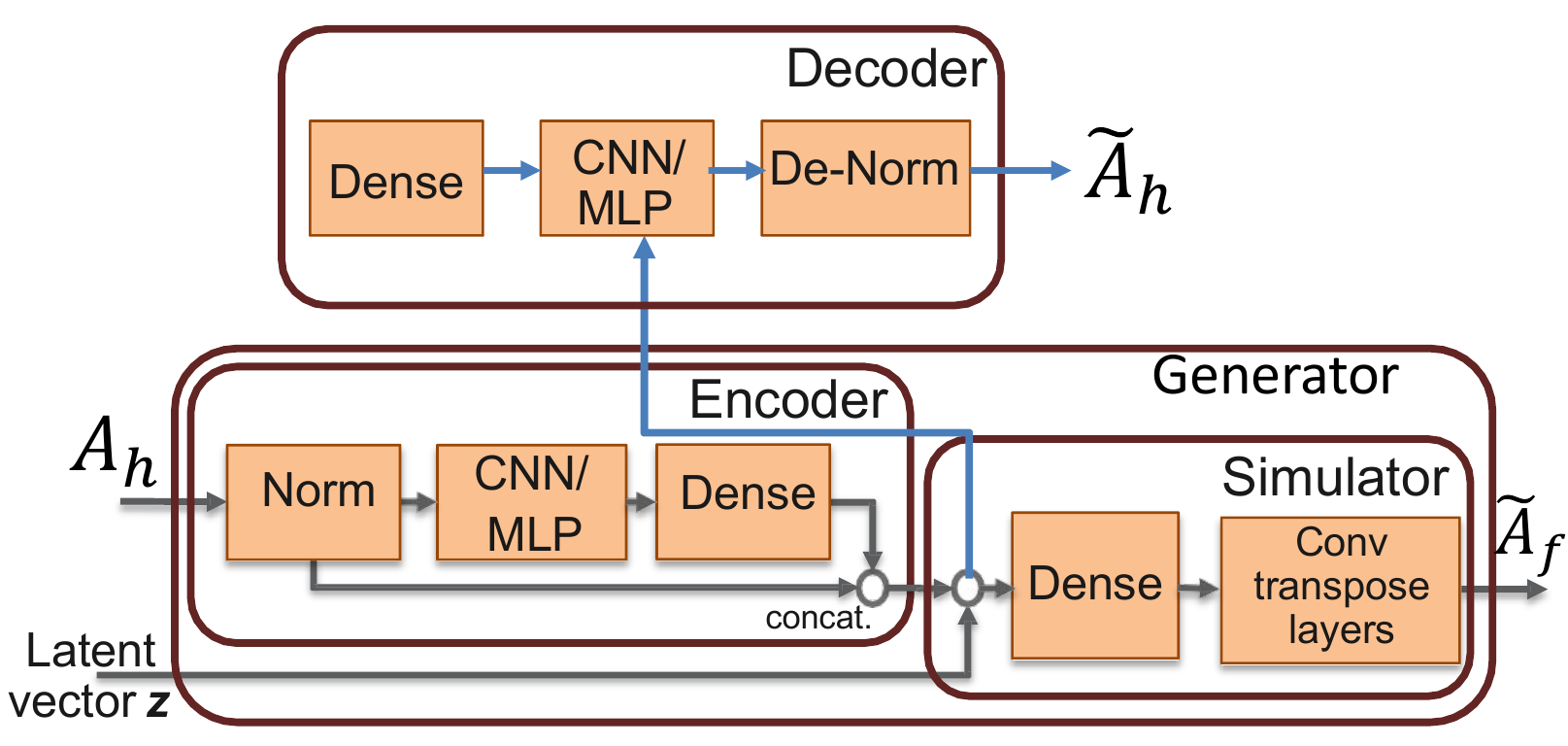}
	\caption{Architecture of the ACGAN generative model for portfolio analysis.}
	\label{fig:acgan_structure}
\end{figure}

\section{Autoencoding Conditional GAN (ACGAN) for Portfolio Analysis}
\subsection{Proposed Methodology}
The proposed ACGAN has the same discriminator structure as the CGAN. However, it contains an extra \textit{decoder} in the generator as shown in Figure~\ref{fig:acgan_structure}. And therefore we call the conditioner as an \textit{encoder} in the ACGAN context.

We use an \textit{encoding} deep-neural network $E$ to learn the features that can help the generator trick the discriminator and can find the internal information itself; and a \textit{decoding} deep-neural network $F$ to reconstruct the historical series so as to force the encoder to do so. Formally the encoding and decoding models reconstruct the historical matrix by
$$
\begin{aligned}
	\by = E(\bA_h), \gap \widetilde{\bA}_h =F(\by).
\end{aligned}
$$ 
This process is known as the \textit{autoencoding}, hence the name autoencoding conditional GAN (ACGAN). In a non-GAN context, the autoencoder is typically done by matrix decomposition or nonnegative matrix factorization via alternative least squares or Bayesian inference \citep{lee1999learning, lu2021numerical, lu2022matrix, lu2022flexible}. Since we need to use the encoding part of the autoencoder to help ``cheat" the discriminator as well, we here use deep-neural network instead. Formally the process is described by the following optimization:
\begin{equation}\label{equation:acgan_propose_losses}
\begin{aligned}
&\mathop{\max}_{D} &\,&\mathbb{E}_{\bx\sim p(\text{data})}
\bigg\{D(\bx) - \mathbb{E}_{\bz\sim p(\bz)}\big[D(G(\bz,\bx_h))\big] \bigg\}
-\\
&& &\,\,\, \lambda_1\mathbb{E}_{\overline{\bx}\sim p(\epsilon\, \text{data} + (1-\epsilon) G(\bz))}
\big[ ||\nabla_{\overline{\bx}} D(\overline{\bx}) ||_2 -1 \big]^2;\\
& \mathop{\max}_{G, E, F} &\, &\mathbb{E}_{\bx\sim p(\text{data})} \bigg\{\mathbb{E}_{\bz\sim p(\bz)}\big[ D(G(\bz, \bx_h))  \big]-\\
&&& \,\,\, \textcolor{blue}{\lambda_2 \,f\big(\underbrace{F(E(\bx_h))}_{\widetilde{\bx}_h},\,\, \bx_h\big)}\bigg\}
,\\
\end{aligned}
\end{equation}
where $f(\cdot)$ denotes the loss function. In our work, we apply the mean squared error (MSE) as the loss function. The parameter $\lambda_2$ controls how large the penalization by the autoencoder, and we call the term \textit{autoencoding penalty (AP)}.
In the original CGAN scenario, the conditioner is used to extract features that can help the generator to cheat the discriminator; however, it may lose some important information that captures the features of the market trend. The ACGAN then finds a balance between cheating the discriminator and retaining its market information.

\paragraph{Other Extension} In our work, we apply to find small discrepancy between $F(E(\bx_h))$ and $\bx_h$ in Eq.~\eqref{equation:acgan_propose_losses} so that the encoder can keep the original information of $\bx_h$ as much as possible. If one believes the encoder should keep information to some other features, say $\by_h$ (e.g., the fat-tail, mean, skewness of the data), the ACGAN can be extended to the general \textit{encoding CGAN (ECGAN)} that penalizes the following loss:
\begin{equation}
f\big(F(E(\bx_h)),\,\, \textcolor{blue}{\by_h}\big).
\end{equation}

\paragraph{Discriminator} The discriminator shown in Figure~\ref{fig:discriminator_cgan} (for both CGAN and ACGAN) takes as input either the real data matrix $\bA=[\bA_h, \bA_f]\in\real^{N\times w}$ or the synthetic data matrix $\widetilde{\bA}=[\bA_h, \widetilde{\bA}_f]\in \real^{N\times w}$.

\subsection{Data Normalization}
Following \citet{mariani2019pagan}, we consider the \textit{adjusted closing price} $\bp\in \real^w$ series for each asset. Given the frame window of $w=h+f$ days ($h$ for the historical length, $f$ for the future length. The historical series is denoted by $\bp_{1:h}\in \real^h$, and the real future series can be obtained by $\bp_{h+1:w}\in \real^f$), we unit-normalize the price series $\bp$ for each asset to fill in the range $[-1,1]$ for the initial $h$ days. In practice, the unit-normalization can be done by 3-sigma normalization: given the mean $\mu$ and standard deviation $\sigma$ of $\bp_{1:h}\in \real^h$, the normalization is done by
\begin{equation}
\widetilde{\bp} = \frac{\bp-\mu}{3\sigma}.
\end{equation}
This normalization procedure can help us to expose the neural networks values limited within a suitable range that removes price-variability over multiple assets within the specified window.

After generating the surrogate future series $\widetilde{\bp}_{h+1:w}$, we apply again a de-normalization procedure:
\begin{equation}\label{equation:denormalization-acgan}
\widehat{\bp}_{h+1:w} = \widetilde{\bp}_{h+1:w} \times 3\sigma +\mu.
\end{equation}

\begin{table*}
	\parbox{.47\linewidth}{
		\centering
\scriptsize
\begin{tabular}{l|lllll}
	\hline
	& Ticker & Type & Sector & Company & Curr. \\ \hline
	\parbox[t]{0.3mm}{\multirow{10}{*}{\rotatebox[origin=c]{90}{US Region}  }} 
	& GOOG   & Share & IT               & Alphabet                 & USD   \\
	& MSFT   & Share & IT               & Microsoft                & USD   \\
	& PFE    & Share & Healthcare       & Pfizer                   & USD   \\
	& HES    & Share & Energy           & Hess                     & USD   \\
	& XOM    & Share & Energy           & Exxon Mobil              & USD   \\
	& KR     & Share & Consumer staples & The Kroger               & USD   \\
	& WBA    & Share & Consumer staples & Walgreens  Alliance & USD   \\
	& IYY    & ETF   & Dow Jones        & iShares Dow Jones        & USD   \\
	& IYR    & ETF   & Real estate      & iShares US Real Estate   & USD   \\
	& SHY      & ETF   & US treasury bond & iShares Treasury Bond    & USD   \\
	\hline
	\hline 
	\parbox[t]{0.3mm}{\multirow{10}{*}{\rotatebox[origin=c]{90}{EU Region}  }} 
	& $~^\wedge$FCHI  & Index & French market    & CAC 40                   & EUR   \\
	& $~^\wedge$GDAXI & Index & German market    & DAX                      & EUR   \\
	& BMW.DE   & Share & Automotive       & BMW                      & EUR   \\
	& VOW3.DE  & Share & Automotive       & Volkswagen               & EUR   \\
	& SOI.PA   & Share & Industrials      & Soitec S.A.              & EUR   \\
	& VK.PA    & Share & Industrials      & Vallourec S.A.           & EUR   \\
	& BAS.DE   & Share & Basic materials  & BASF SE                  & EUR   \\
	& SAP.DE   & Share & Technology       & SAP SE                   & EUR   \\
	& DTE.DE   & Share & Technology       & Deutsche Telekom AG      & EUR   \\
	& BAYN.DE  & Share & Healthcare       & Bayer AG & EUR   \\
	\hline
\end{tabular}
\caption{Summary of the underlying portfolios in the US and EU markets, 10 assets for each market respectively. In each region, we include assets from various sectors to favor a somehow sector-neutral strategy.}
\label{table:us_eu_data_summary}		
	}
	\hfill
	\parbox{.47\linewidth}{
		\centering
	\scriptsize
\begin{tabular}{lll|lll}
	\hline
	Asset &  CGAN &  ACGAN & Asset &  CGAN &  ACGAN \\	\hline 
	GOOG & 9.968 & \textbf{9.972} &  $~^\wedge$FCHI  &\textbf{9.875} &9.838\\ 
	MSFT & \textbf{9.967} & 9.961 &  $~^\wedge$GDAXI  &9.849 &\textbf{9.870}\\ 
	PFE & 9.836 & \textbf{9.851} &  BMW.DE  &9.824 &\textbf{9.867}\\ 
	HES & 9.831 & \textbf{9.871} &  VOW3.DE  &9.732 &\textbf{9.745}\\ 
	XOM & \textbf{9.888} & 9.877 &  SOI.PA  &9.873 &\textbf{9.895}\\ 
	KR & 9.902 & \textbf{9.948} &  VK.PA  &{9.965} &\textbf{9.969}\\ 
	WBA & 9.690 & \textbf{9.750} &  BAS.DE  &\textbf{9.859} &9.854\\ 
	IYY & \textbf{9.948} & 9.945 &  SAP.DE  &9.063 &\textbf{9.138}\\ 
	IYR & 9.833 & \textbf{9.839} &  DTE.DE  &9.879 &\textbf{9.901}\\ 
	SHY & \textbf{9.934} & {9.933} &  BAYN.DE  &9.502 &\textbf{9.513}\\ 
	\hline
	\hline
	GOOG & \textbf{9.958} & 9.954 &  $~^\wedge$FCHI  &\textbf{9.865} &9.864\\ 
	MSFT & 9.940 & \textbf{9.948} &  $~^\wedge$GDAXI  &\textbf{9.866} &9.864\\ 
	PFE & 9.847 & \textbf{9.860} &  BMW.DE  &9.837 &\textbf{9.873}\\ 
	HES & \textbf{9.848} & 9.840 &  VOW3.DE  &9.721 &\textbf{9.765}\\ 
	XOM & \textbf{9.873} & 9.862 &  SOI.PA  &9.914 &\textbf{9.927}\\ 
	KR & 9.921 & \textbf{9.928} &  VK.PA  &{9.976} &\textbf{9.977}\\ 
	WBA & 9.495 & \textbf{9.659} &  BAS.DE  &{9.824} &\textbf{9.825}\\ 
	IYY & 9.930 & \textbf{9.939} &  SAP.DE  &9.364 &\textbf{9.466}\\ 
	IYR & 9.810 & \textbf{9.833} &  DTE.DE  &{9.889} &\textbf{9.890}\\ 
	SHY & 9.904 & \textbf{9.916} &  BAYN.DE  &{9.615} &\textbf{9.616}\\ 
	\hline
\end{tabular}
\caption{Mean Pearson correlation between the true series and the generated series. The fake series are generated by the generator at the 100-th epoch (upper table) and the 1,000-th epoch (lower table). The values are multiplied by 10 for clarity.}
\label{table:correlation_acgan-cgan}
	}
\end{table*}

\subsection{Statistical Properties of Financial Time Series}\label{section:statistica_property_financial_tim}
We have several statistical properties (stylized facts) of financial time series \citep{muller1997volatilities, cont2001empirical, chakraborti2011econophysics, takahashi2019modeling}. Given the price series $p_t$ of an asset at time $t$, the log return of price can be obtained by 
$$
r_t= \log p_{t+1} - \log p_t.
$$
We then review a few facts on the return series.
\paragraph{Linear Unpredictability}
The most important fundamental property of the return series is its linear unpredictability (a.k.a., absence of autocorrelations) that is quantified by its diminishing autocorrelation of the return series:
$$
\text{Corr}(r_t, r_{t+k}) =
\frac{\mathbb{E} [(r_t-\mu)(r_{t+k}-\mu)] }{\sigma^2} \approx 0, \forall\,\, k\geq 1,
$$
where $\mu, \sigma$ are the mean and standard deviation of the return series respectively.

\paragraph{Fat-Tailed Distribution}
The probability distribution function of the return series $p(r)$ is empirically known to have a power-law decay \footnote{The distribution is known as the Zipf distribution or Zeta distribution.} in the tails:
$$
p(r) \propto r^{-\alpha}.
$$
Empirically, the exponent $\alpha$ ranges $3\leq \alpha\leq  5$.

\paragraph{Leverage Effect}
The leverage effect means that there is a negative correlation between past price return and future volatility \citep{bouchaud2001leverage}. In other words, if the market declines significantly in the past price,
the future volatility will increase; while if the market increases significantly in the past price, the future volatility will decrease.
The leverage effect is quantified by the following lead-lag correlation function
$$
L(k) = \frac{\mathbb{E}[r_t |r_{t+k}|^2] - 
\mathbb{E}[	r_t] \cdot \mathbb{E}[|r_t|^2]}{\mathbb{E}[|r_t|^2]^2}.
$$
\citet{bouchaud2001leverage} show that $L(k)$ has a negative value for $1\leq k \leq 10$ and the distribution follows the exponential decay.

\paragraph{Coarse-Fine Volatility Correlation} The coarse-fine volatility correlation is a multi-time-scale analysis of volatility \citep{muller1997volatilities, rydberg2000realistic}. Define the coarse volatility $\nu_c^\tau $ and the fine volatility $\nu_f^\tau$ as 
$$
\nu_c^\tau(t) = \bigg|\sum_{i=1}^{\tau} r_{t-i} \bigg|, \gap \nu_f^\tau=\sum_{i=1}^{\tau} |r_{t-i} |,
$$
where the coarse volatility is the absolute value of the movement in $\tau$ days, and the fine volatility is the sum of absolute return in $\tau$ days. Then we calculate the correlation between the current fine volatility and $k$-th lagged coarse volatility:
$$
\rho_{cf}^\tau(k) = \text{Corr}(\nu_c^\tau (t+k), \nu_f^\tau(t)).
$$
Then there exists the negative asymmetry (especially when $k$ is small) of the lead-lag correlation quantified by the following difference
$$
\Delta \rho_{cf}^\tau (k) =  \rho_{cf}^\tau(k) - \rho_{cf}^\tau(-k) <0.
$$
The fact that $\Delta \rho_{cf}^\tau (k)$ is negative means that fine volatility can predict coarse volatility.

\begin{figure}[h]
	\centering  
	\subfigtopskip=2pt 
	\subfigbottomskip=2pt 
	\subfigcapskip=-5pt 
	\subfigure{\includegraphics[width=0.425\textwidth]{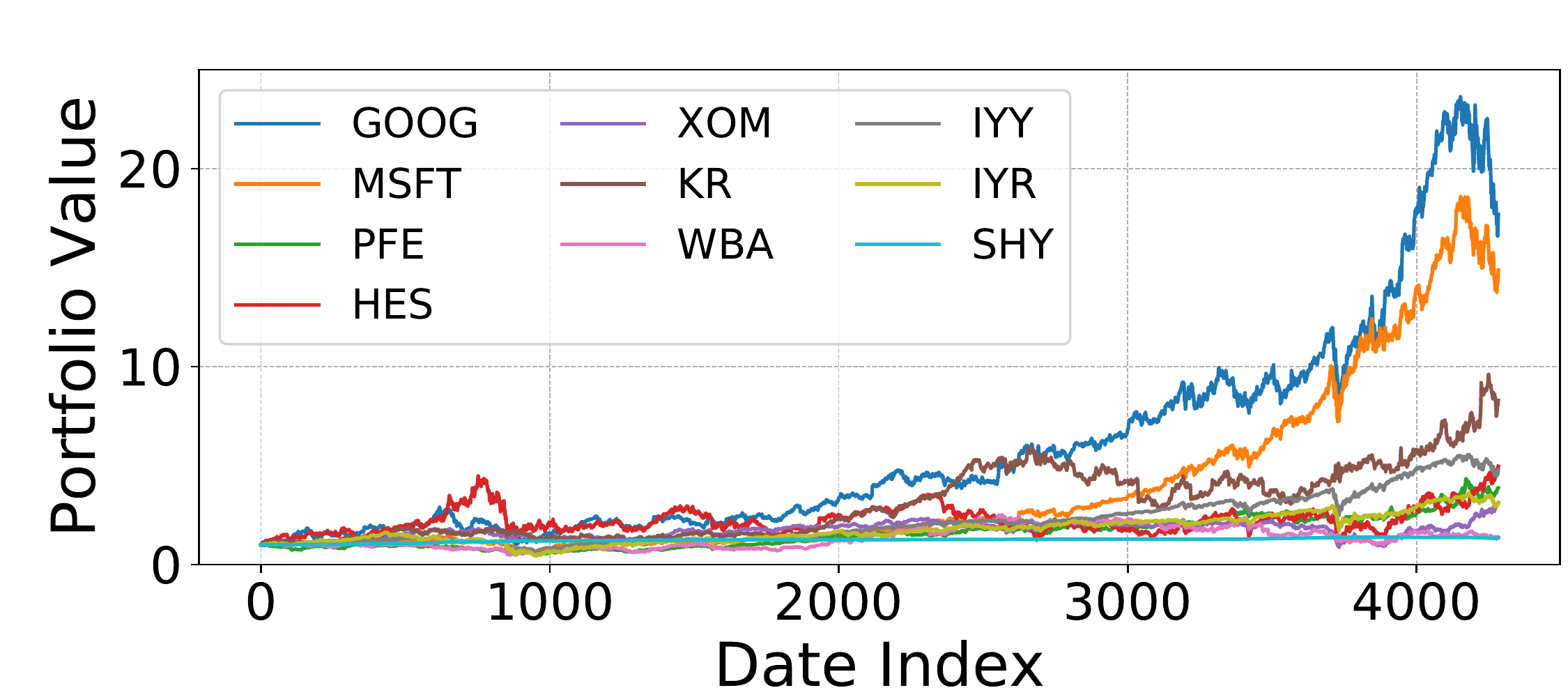} \label{fig:acgan_dataset_us}}
	\subfigure{\includegraphics[width=0.425\textwidth]{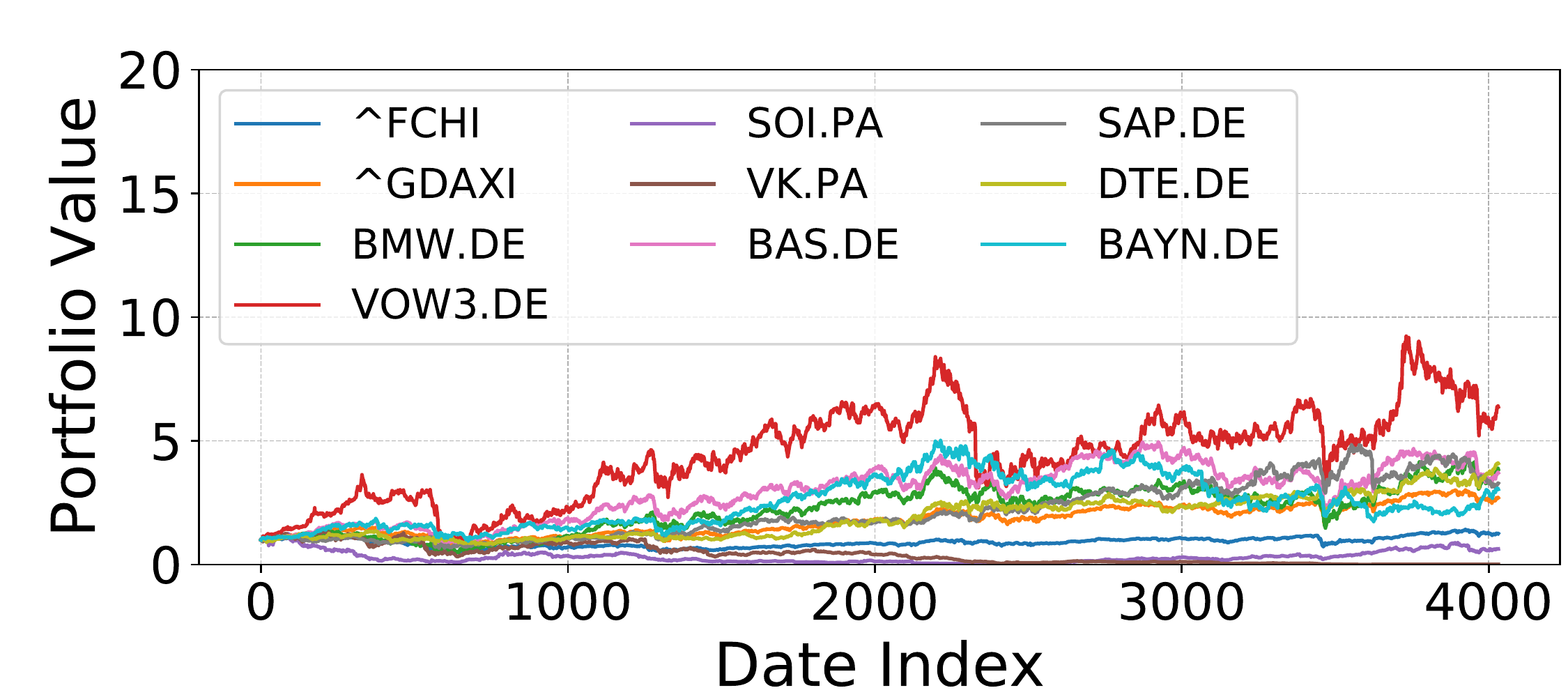} \label{fig:acgan_dataset_eu}}
	\caption{Different portfolios for the US (upper) and EU (lower) markets with a unit initial value.}
	\label{fig:acgan_dataset_us_eu}
\end{figure}

\section{Experiments}\label{section:ader_experiments}

\begin{algorithm}[!htb]
	\caption{Training and testing process for the ACGAN and CGAN models.} 
	\label{alg:acgan_training}  
	\begin{algorithmic}[1] 
		\STATE {\bfseries General Input:} Choose parameters $w=h+f$; number of assets $N$; number of epoches $T$; latent dimension $m$;
		\STATE {\bfseries Training Input: } Training data matrix $\bM\in \real^{N\times D}$;
		\STATE Decide index set $\mathcal{S}_1=\{1,2,\ldots, D-w+1\}$ and \textbf{draw without replacement};
		\FOR{$t=1$ to $T$}
\FOR{$i$ $\in$ $\text{\textcolor{blue}{random}}(\mathcal{S}_1)$} 
\STATE $\bA = [\bA_h, \bA_f]=\bM[:,i:i+w-1]\in \real^{N\times w}$;
\STATE Randomly sample latent vector $\bz\in \real^m$;
\STATE Backpropatation for generator in Eq.~\eqref{equation:acgan_propose_losses} or \eqref{equation:cgan_losses};
\STATE Generate surrogate $\widetilde{\bA}_f = G( \bz,\bA_h)\in \real^{N\times f}$;
\STATE Backpropatation for discriminator in Eq.~\eqref{equation:acgan_propose_losses} or \eqref{equation:cgan_losses};
\ENDFOR
		\ENDFOR
		\STATE {\bfseries Inference Input: } Testing data matrix  $\bX\in \real^{N\times K}$;
		\STATE {\bfseries Inference Output: } Testing data matrix $\bY\in \real^{N\times K}$;
		\STATE Decide index set $\mathcal{S}_2=\{\textcolor{blue}{h+1},h+f+1,\ldots\}$;
		\STATE Copy the first $h$ days data $\bY[:,1:h]=\bX[:,1:h]$;
		\FOR{$i$ $\in$ $\text{\textcolor{blue}{ordered}}(\mathcal{S}_2)$} 
		\STATE $\bA = [\bA_h, \bA_f]=\bX[:,i:i+w-1]\in \real^{N\times w}$;
		\STATE Randomly sample latent vector $\bz\in \real^m$;
		\STATE Generate $\bY[:,i:i+f-1] = G(\bz,\bA_h)\in \real^{N\times f}$ with de-normalization in Eq.~\eqref{equation:denormalization-acgan};
		\ENDFOR
		\STATE Output the synthetic series $\bY$;
	\end{algorithmic} 
\end{algorithm}

\begin{figure*}[h]
\centering  
\subfigtopskip=2pt 
\subfigbottomskip=2pt 
\subfigcapskip=-2pt 
\subfigure{\includegraphics[width=1\textwidth]{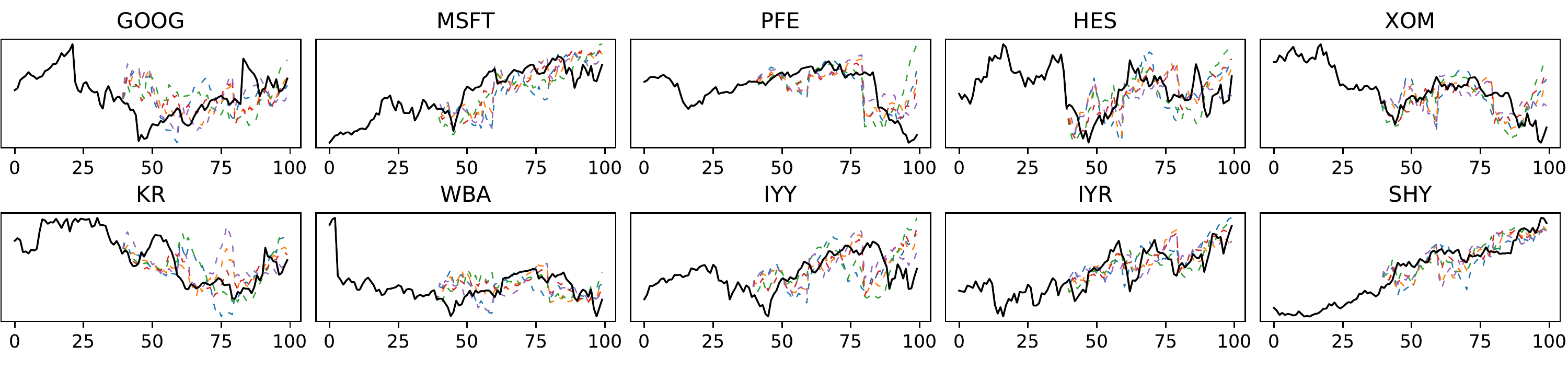}}
\subfigure{\includegraphics[width=1\textwidth]{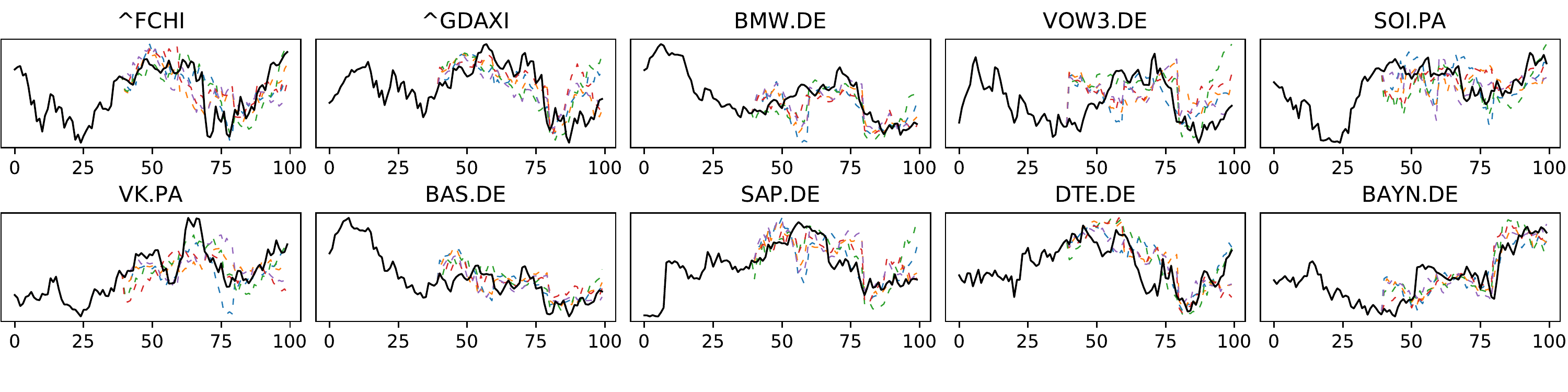}}
\caption{Actual price trend (black solid line) of the US assets (upper two rows) and the EU assets (lower two rows)  for the first 100 trading days in the test set, and five representative simulations generated by ACGAN (colored dashed lines). The simulated price series of the whole period (800 trading days) and simulated price series of the CGAN model (100 and all trading days) can be found in Figure~\ref{fig:diversity_us_eu_acgan_800}, \ref{fig:diversity_us_eu_cgan}, and \ref{fig:diversity_us_eu_cgan_800} respectively.}
\label{fig:diversity_us_eu_ecgan}
\end{figure*}

To evaluate the strategy and demonstrate the main advantages of the proposed ACGAN method, 
we conduct experiments with different analysis tasks; datasets from different geopolitical markets including the US and the European (EU) markets, and various industrial segments including Healthcare, Automotive, Energy and so on.
We obtain public available data from Yahoo Finance \footnote{\url{https://finance.yahoo.com/}.}. 

For the US market, we obtain data for a 17-year period, i.e., 2005-05-24 to 2022-05-27, where the data between  2005-05-24 and 2019-03-28 is considered training data; 
while data between 2019-03-28 and 2022-05-27 is taken as the test data (800 trading days).
For the EU market, we obtain data for a 16-year period, i.e., 2006-07-18 to 2022-06-07, where the data between 2006-07-18 and 2019-04-09 is considered training data;
while data between 2019-04-10 and 2022-06-07 is taken as the test data (800 trading days). 
The underlying portfolios are summarized in Table~\ref{table:us_eu_data_summary}:
\begin{itemize}
\item \textit{US market}: 10 assets of US companies from different industrial segments, i.e., Alphabet and Microsoft (from IT sector), Pfizer (from Healthcare sector), Hess and Exxon Mobil (from Energy sector), Kroger and Walgreens Alliance (from Consumer staples sector), and three ETFs (IYY, IYR, SHY).
\item \textit{EU market}: 10 portfolios of EU companies from different industrial segments, i.e., BMW and Volkswagen (from Automotive sector), Soitec and Vallourec (from Industrials sector), BASF (from Basic materials sector), SAP and Deutsche Telekom (from Technology sector), Bayer (from Healthcare sector), and two indices, $~^\wedge$FCHI and $~^\wedge$GDAXI, that track the German and French stock markets respectively. 
\end{itemize}
The specific time periods and assets are chosen by following the four criteria. \textit{1). Data availability}: we want to cover as longer period as possible; the periods are selected to make all the assets have same frame length; \textit{2). Data diversity}: in each market, we include firms from different segments so that the end strategies are somewhat sector-neutral; \textit{3). Currency homogeneity}: in each marketplace, the currencies are the same; \textit{4). Data correctness}: given the Yahoo Finance data source, we only include the data that do not have NaN values. Figure~\ref{fig:acgan_dataset_us_eu} shows the series of different assets where we initialize each portfolio with a unitary value for clarity.


In all scenarios, same parameter initialization is adopted when conducting different tasks. 
We compare the results in terms of series correlation and performance of portfolio allocation. In a wide range of scenarios across various tasks, ACGAN improves portfolio evaluation, and leads to return-risks performances that are as good or better than the existing Markowitz framework and CGAN methodology.

\paragraph{Hyperparameters}
Network structures for the conditioner (in CGAN), encoder, decoder (in ACGAN), generator, and discriminator (in both CGAN and ACGAN) are provided in Appendix~\ref{appendix:acgan_net_structures}. In all experiments, we train the network with 1,000 epochs. For simplicity, we set the risk-free interest $r_f=0$ to assess the Sharpe ratio evaluations.

\begin{table*}[]
	\begin{tabular}{l|lll|lll}
		\hline
Statistics      & \gap Real (US) & ACGAN (US) & CGAN (US) & \gap Real (EU) & ACGAN (EU) & CGAN (EU) \\ \hline
Autocorrelation & \gap  0.027   &  \textbf{ 0.035 }     &  \textbf{0.036}   & \gap 0.018     & \textbf{ 0.032  }   &   \textbf{0.034}   \\
Fat-tail & \gap  3.819  & \textbf{ 3.477 }    &  3.423      &\gap  4.130   & \textbf{ 3.611  }   &   3.552     \\
Leverage effect  & \gap  -9.328     & \textbf{ -7.254 }    &  -6.384      &\gap  -10.228   &  -7.660    &  \textbf{ -8.815  }   \\
Coarse-fine & \gap  -0.068     & \textbf{ -0.040  } &  -0.011   & \gap -0.074    &  \textbf{-0.033 }   &  \textbf{ -0.038 }    \\
Kurtosis & \gap  11.305     & \textbf{ 22.366  } &  30.019   & \gap 34.002    &   24.144  & \textbf{ 40.297 }  \\
Skewness & \gap  -0.324     & \textbf{ -0.403  } &  0.179   & \gap -1.119    & \textbf{ -0.311 } &  -0.213    \\
		\hline
	\end{tabular}
	\caption{Statistical properties of the real and generated series. Autocorrelation value is the mean of the first 10-th lags of autocorrelation coefficients; Fat-tail value is the fitted powerlaw coefficient $\alpha$, leverage effect value is the mean of the first 10-th lags of leverage effect coefficients; and coarse-fine value is the difference between $\pm$ 1-th order of lead-lag coefficients. In most cases, the coefficients of the ACGAN are closer to the true ones. The Kurtosis of a normal distribution is 3 and for a fat-tail distribution, the value is larger than 3.}
	\label{table:statistics_property-cgan-acgan}
\end{table*}

\begin{figure*}[h]
	\centering  
	\subfigtopskip=2pt 
	\subfigbottomskip=2pt 
	\subfigcapskip=-5pt 
	\subfigure[Linear unpredictability]{\includegraphics[width=0.235\textwidth]{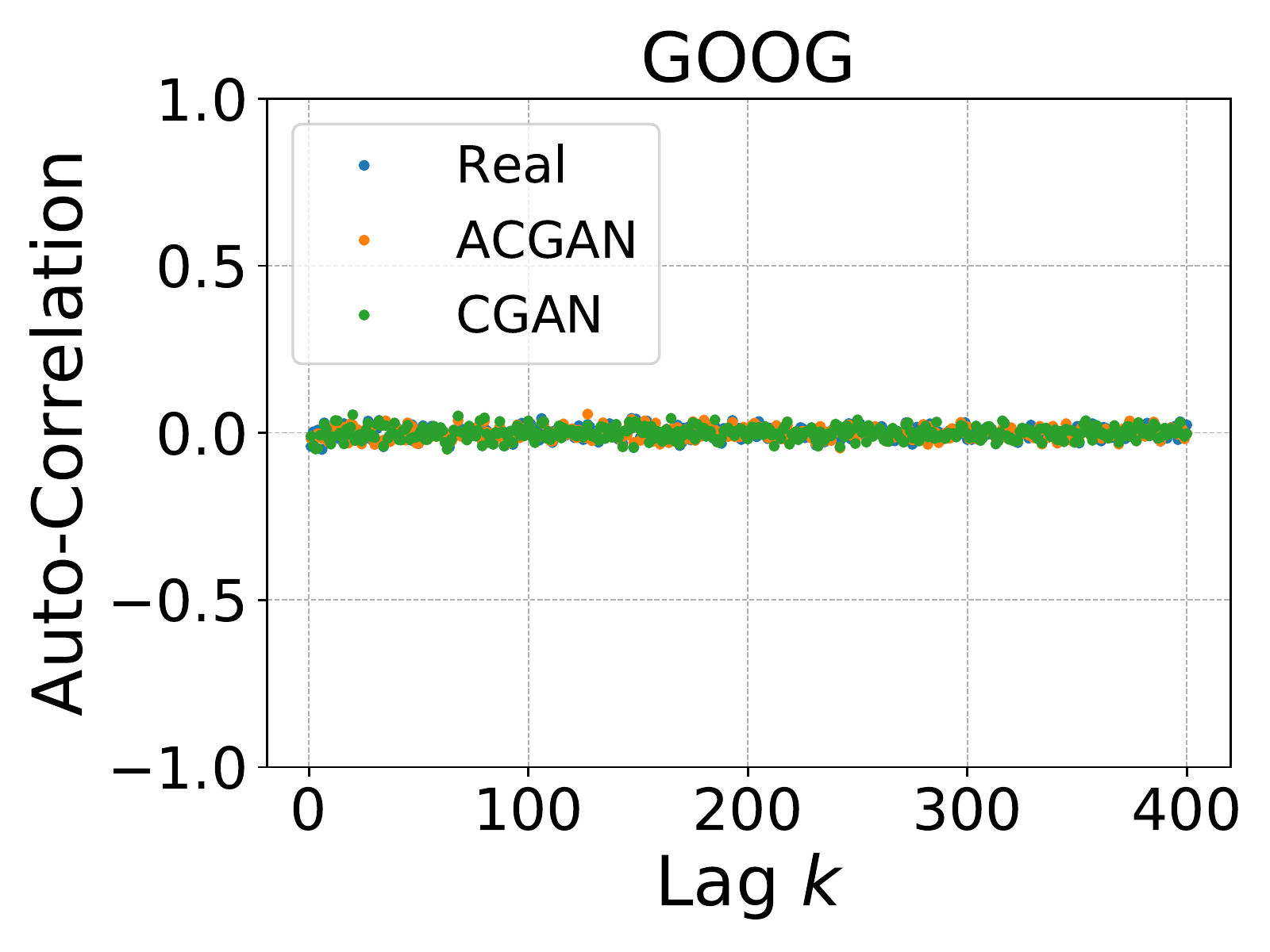} \label{fig:autocorrelation_us_fake_0}}
	\subfigure[Fat-tail distribution]{\includegraphics[width=0.235\textwidth]{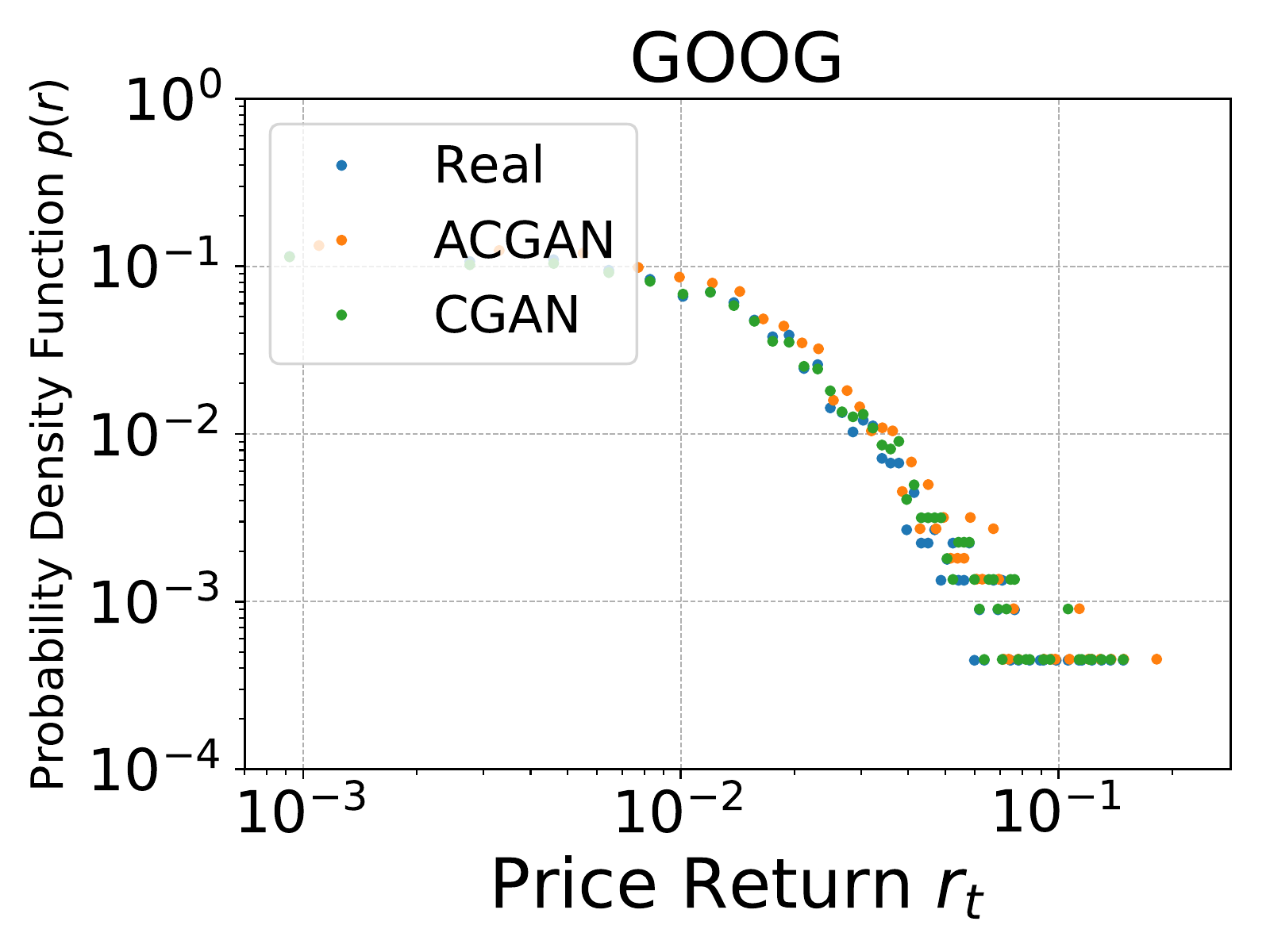} \label{fig:log_positive_distribution_us_fake_ecgan}}
	\subfigure[Leverage effect]{\includegraphics[width=0.235\textwidth]{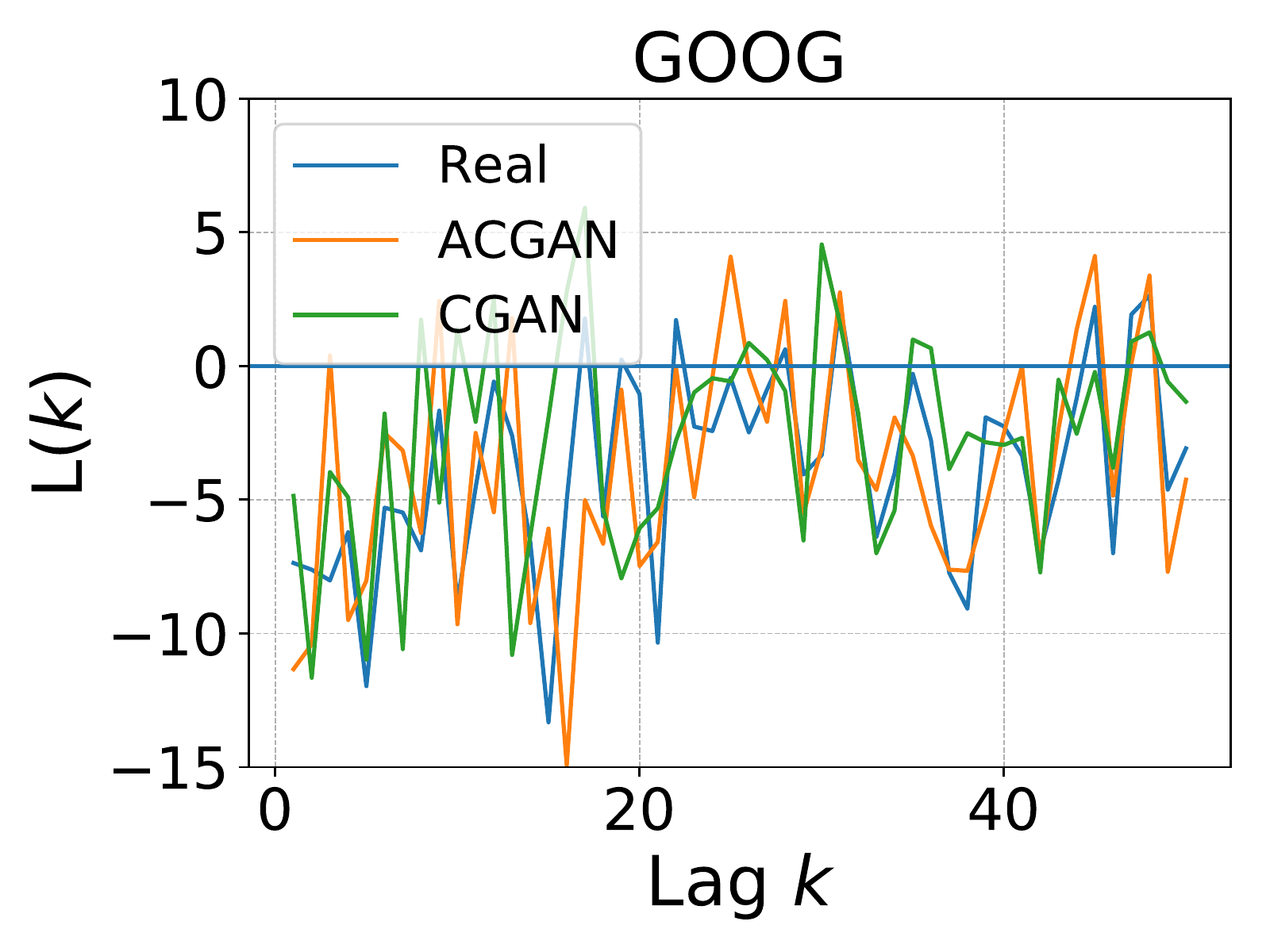} \label{fig:leverage_effect_us_market_0}}
	\subfigure[Coarse-fine volatility correlation]{\includegraphics[width=0.235\textwidth]{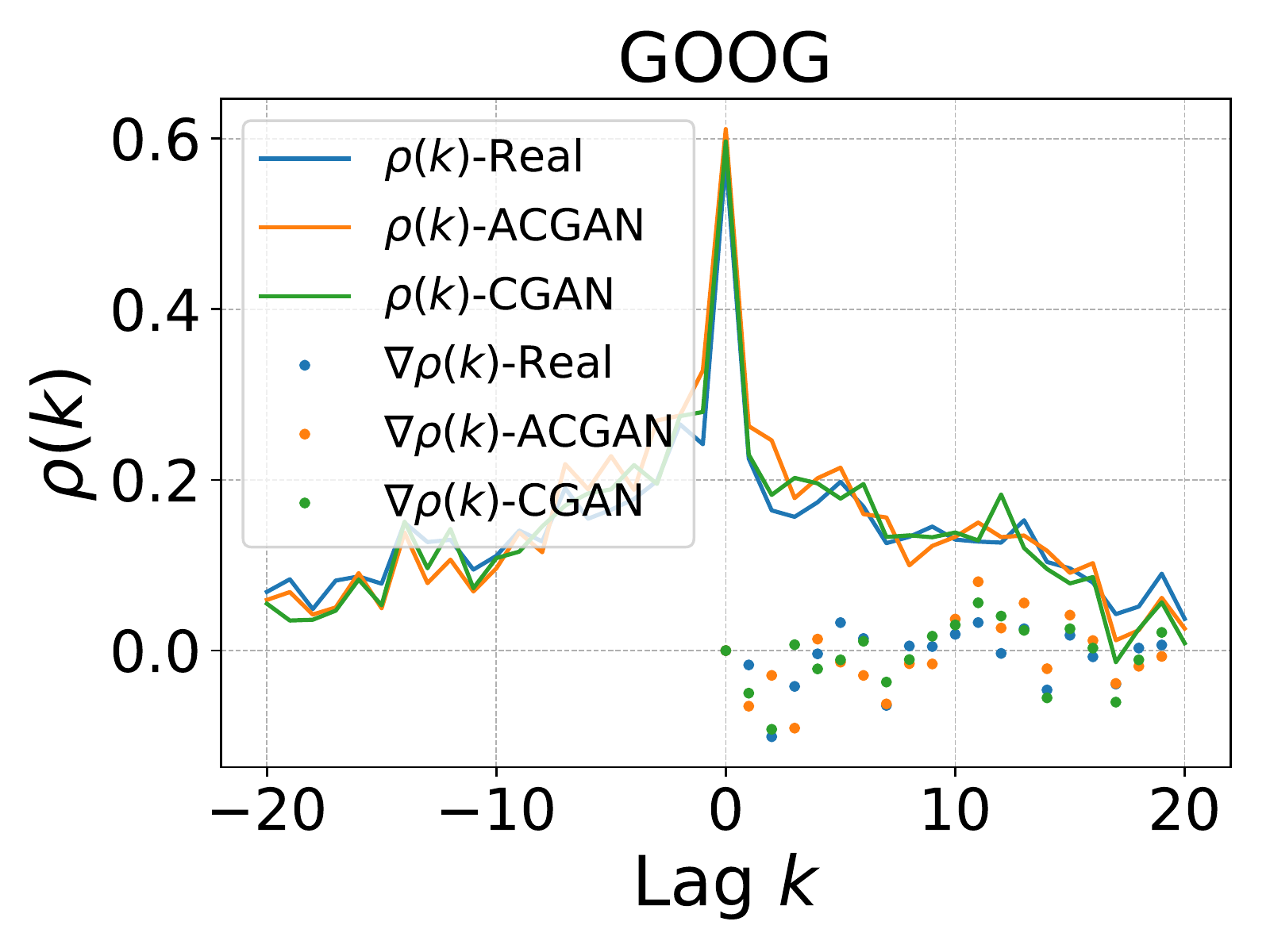} \label{fig:coarse-fine_us_market_0}}
	\caption{Example of autocorrelation, fat-tail distribution, leverage effect, and coarse-fine volatility correlation for one asset.}
	\label{fig:autocorrelation_us_fake_0_and_otherss}
\end{figure*}

\begin{figure*}[!htb]
\centering  
\vspace{-0.35cm} 
\subfigtopskip=2pt 
\subfigbottomskip=2pt 
\subfigcapskip=-5pt 
\subfigure[US, rebalance every 10 days]{\includegraphics[width=0.325\textwidth]{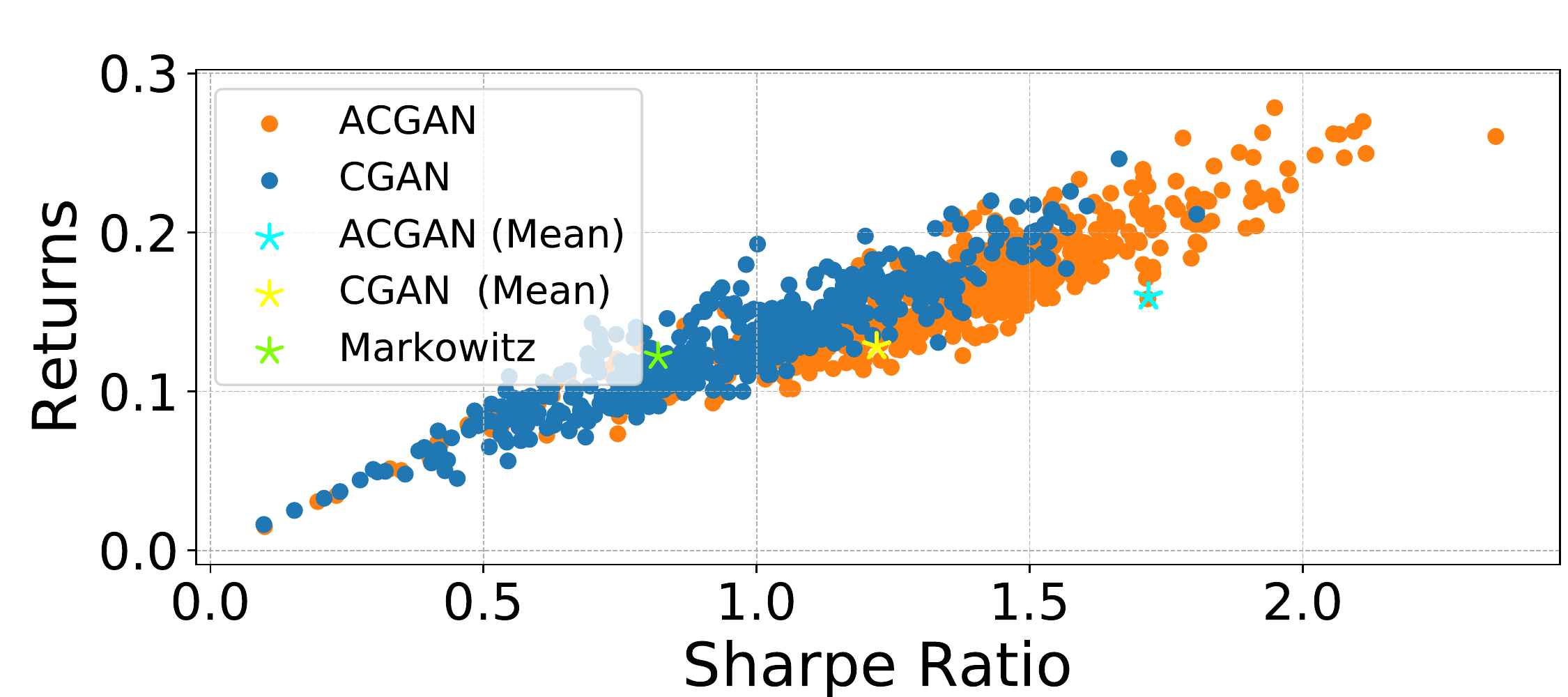} \label{fig:return_sharpe_us1}}
\subfigure[US, rebalance every 15 days]{\includegraphics[width=0.325\textwidth]{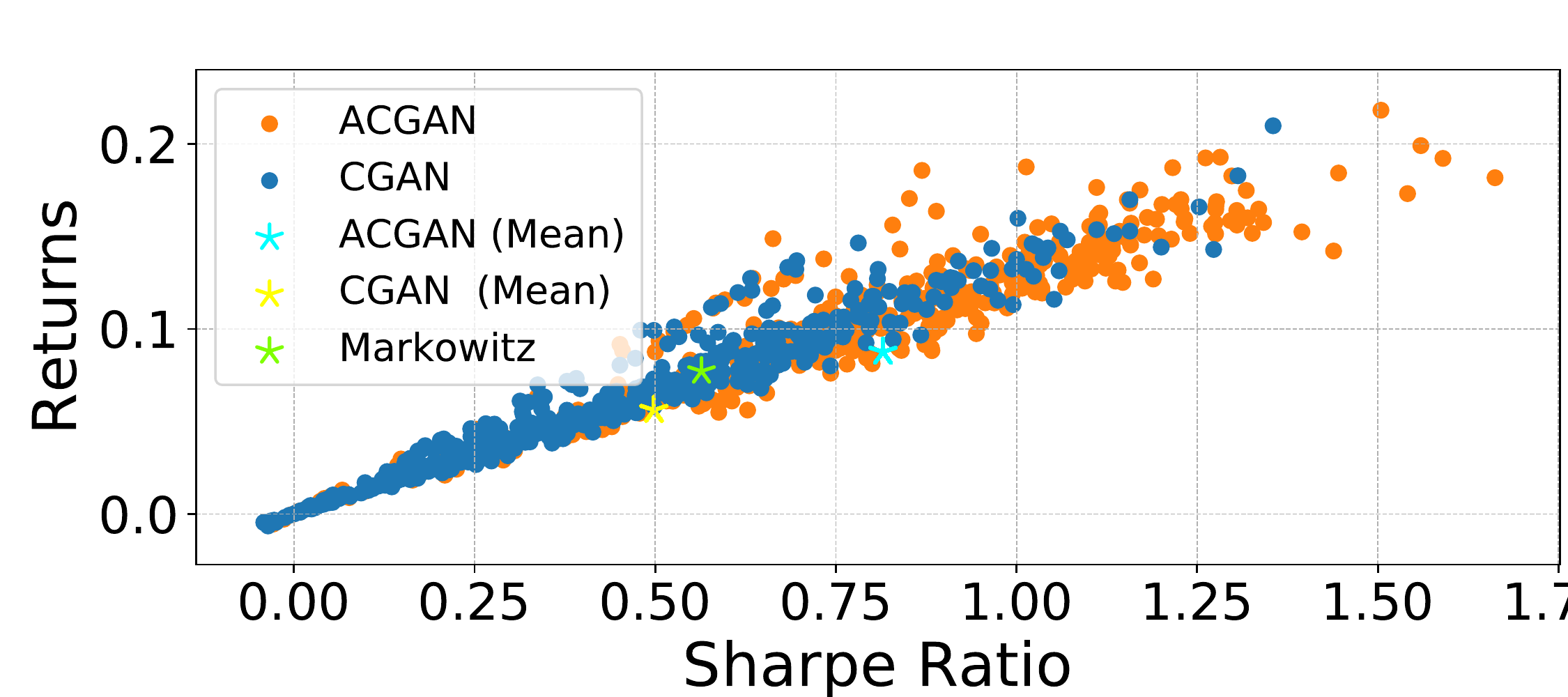} \label{fig:return_sharpe_us2}}
\subfigure[US, rebalance every 20 days]{\includegraphics[width=0.325\textwidth]{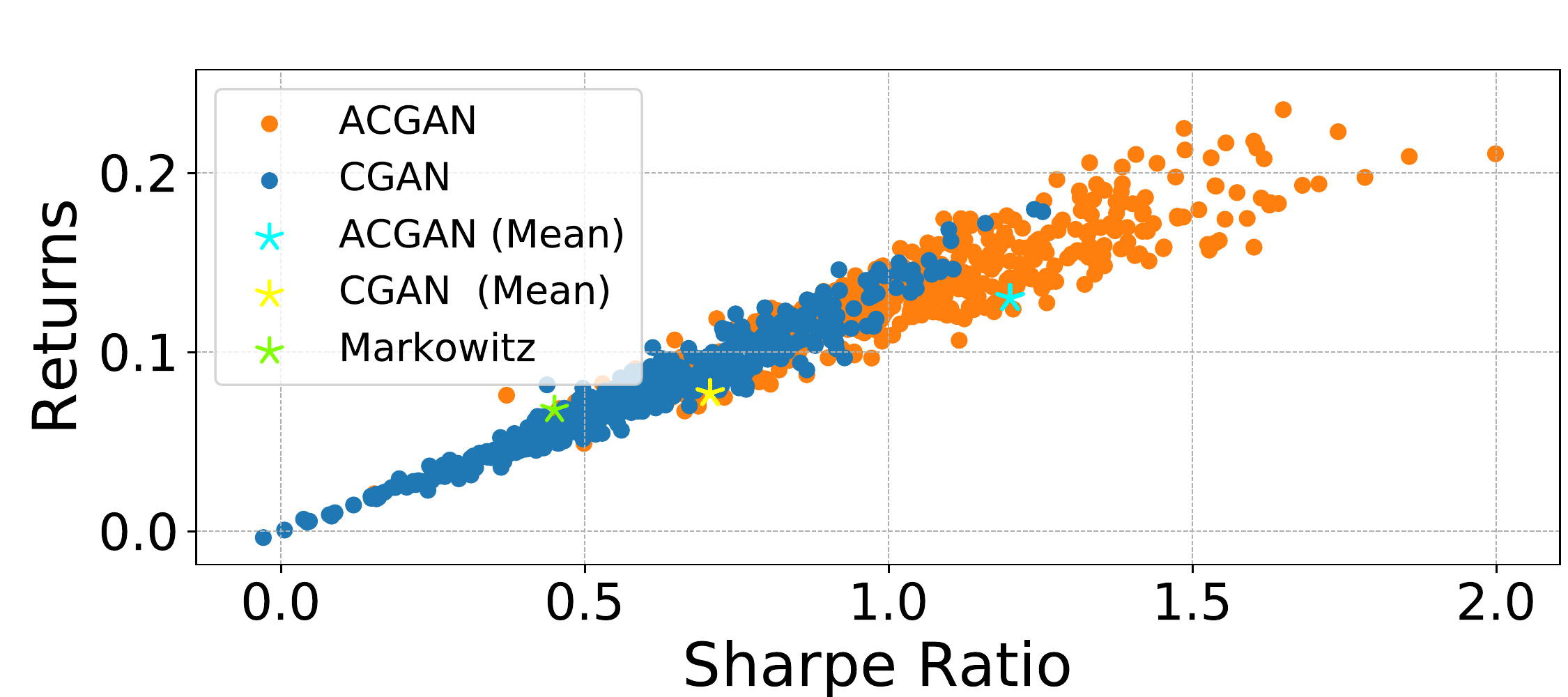} \label{fig:return_sharpe_us3}}
\subfigure[EU, rebalance every 10 days]{\includegraphics[width=0.325\textwidth]{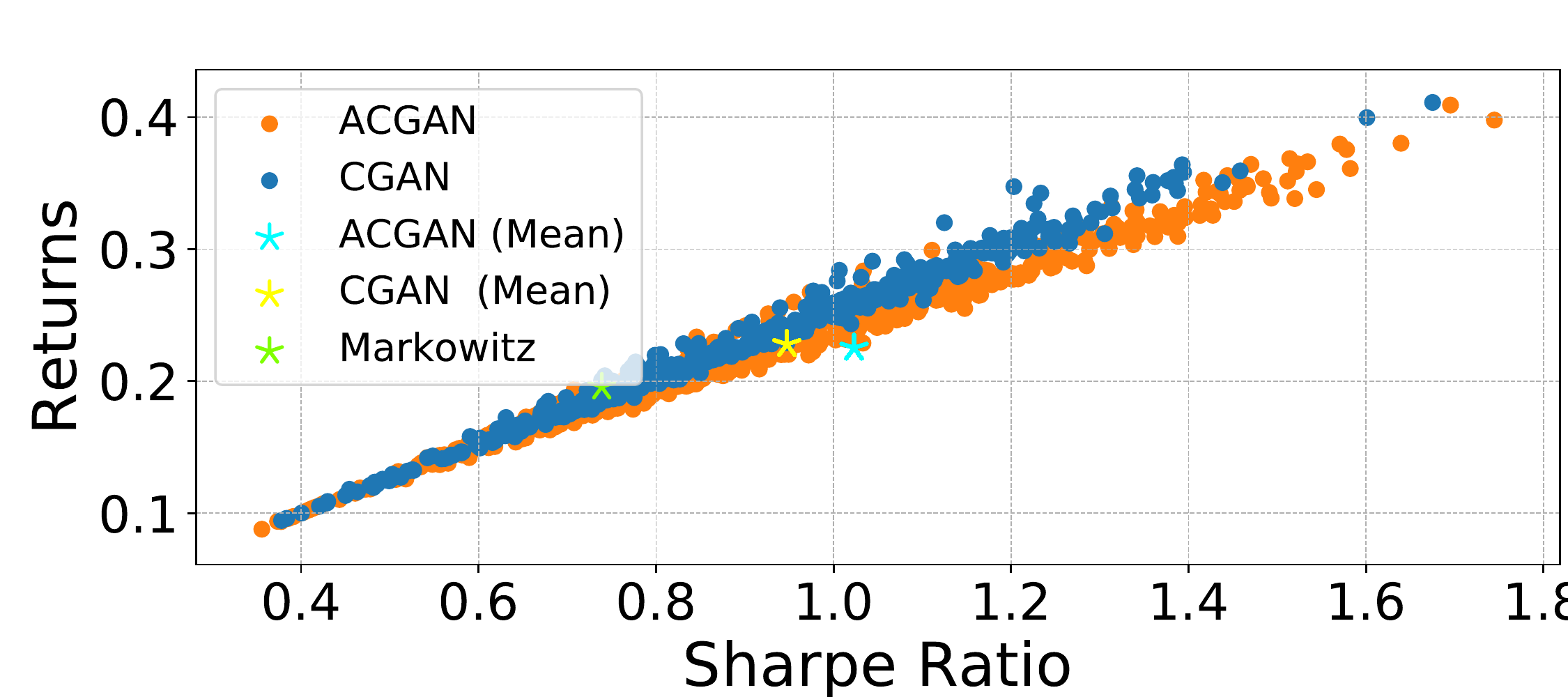} \label{fig:return_sharpe_eu1}}
\subfigure[EU, rebalance every 15 days]{\includegraphics[width=0.325\textwidth]{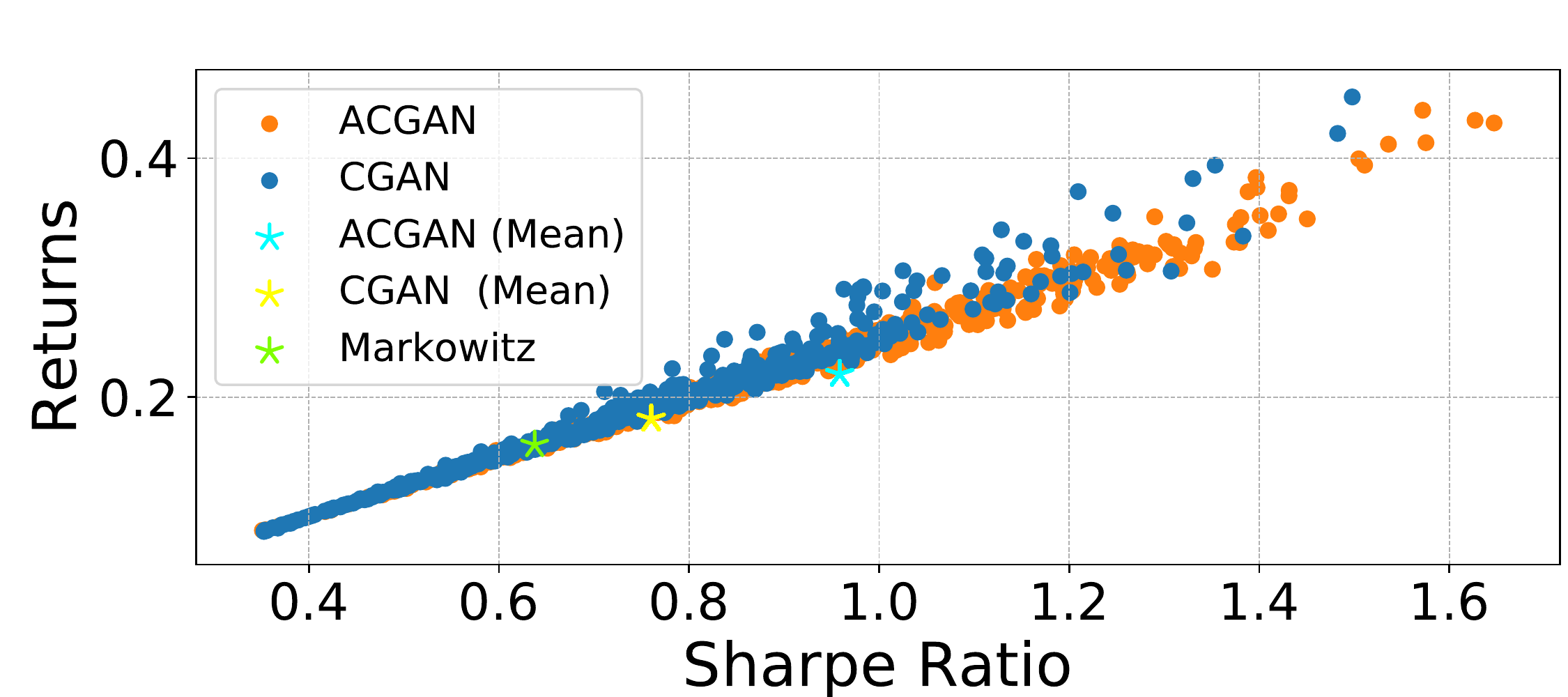} \label{fig:return_sharpe_eu2}}
\subfigure[EU, rebalance every 20 days]{\includegraphics[width=0.325\textwidth]{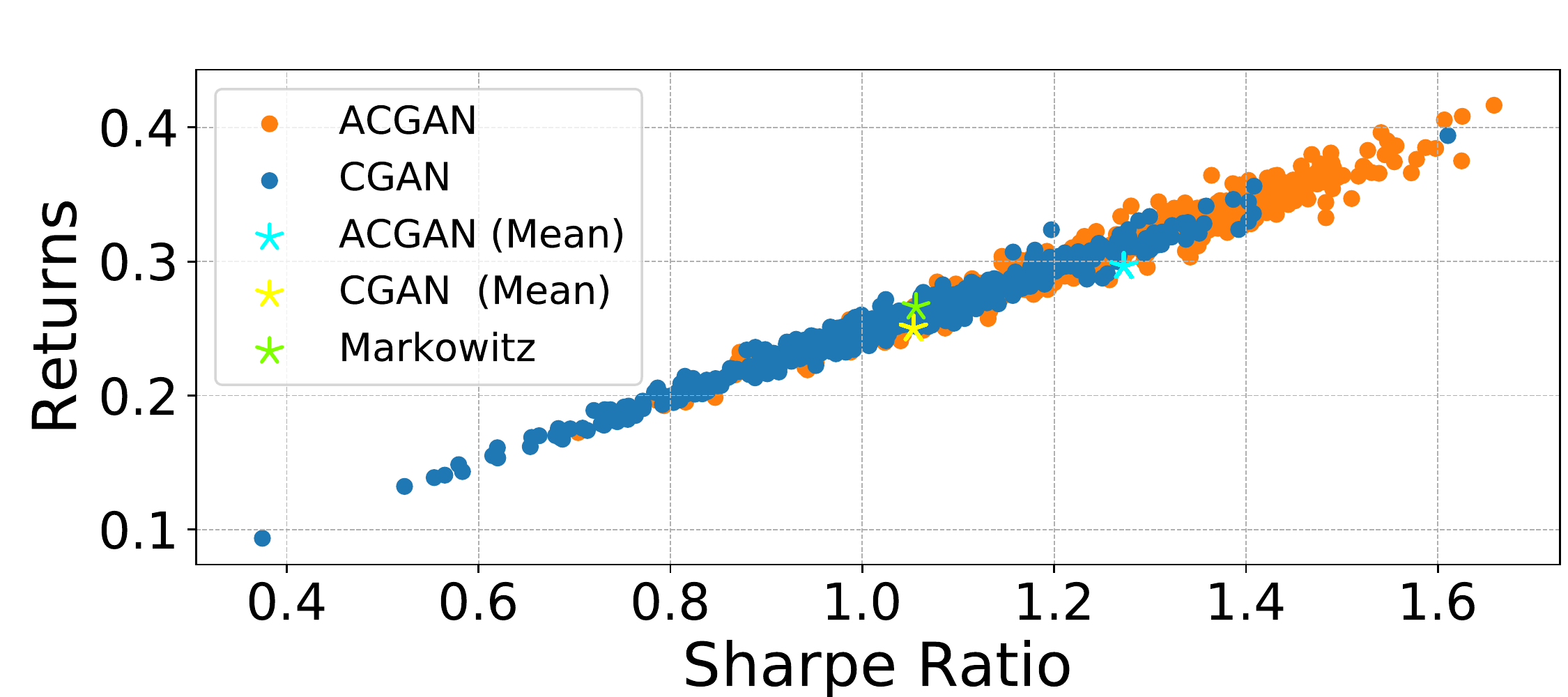} \label{fig:return_sharpe_eu3}}
\caption{(Annual) return-SR measured on the test period by randomly sampling 1000 series.}
\label{fig:risk-sharperation_acgan_compare}
\end{figure*}
\begin{figure*}[!htb]
\centering  
\vspace{-0.35cm} 
\subfigtopskip=2pt 
\subfigbottomskip=2pt 
\subfigcapskip=-5pt 
\subfigure[US, rebalance every 10 days,
Sharpe ratios of ACGAN, CGAN, and Markowitz are \textbf{1.72}, 1.22, and, 0.82 respectively.
]{\includegraphics[width=0.325\textwidth]{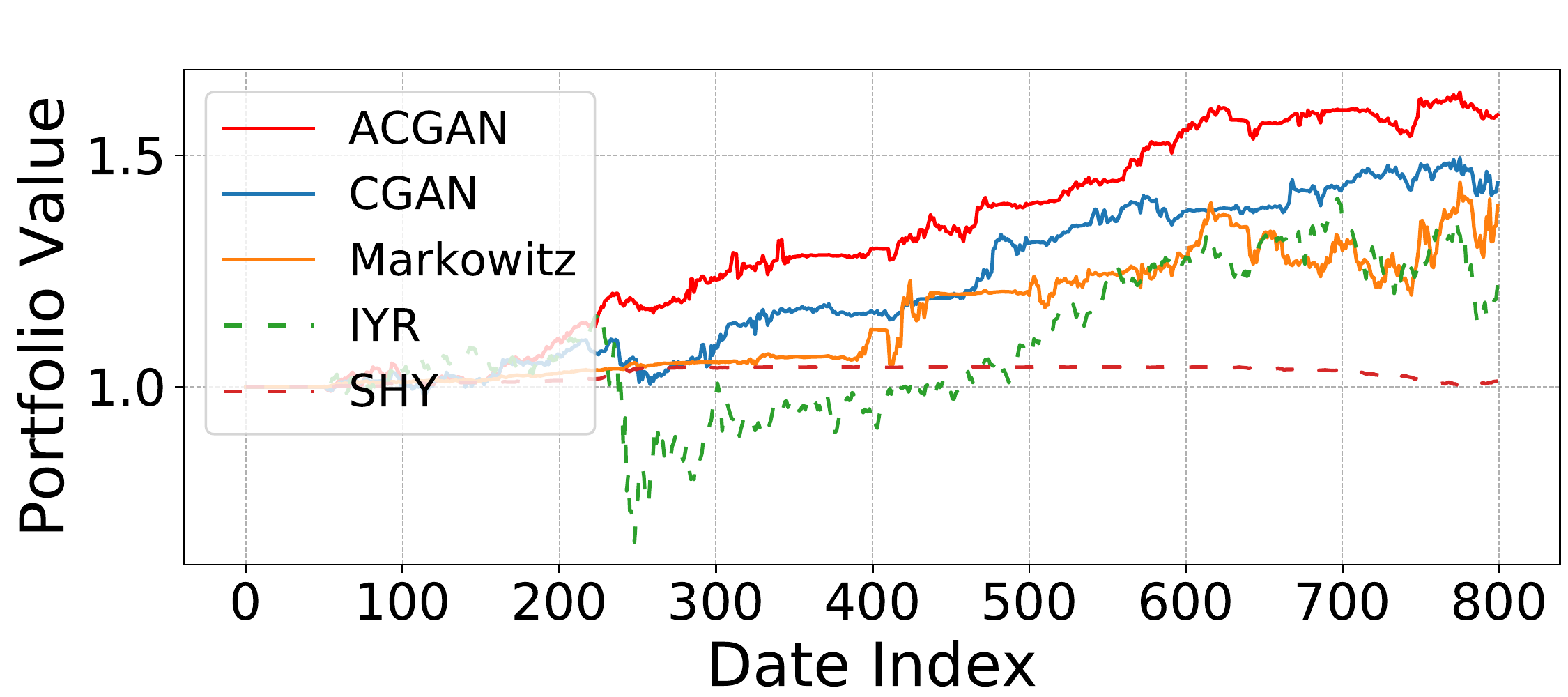} \label{fig:portfolio_value_us1}}
\subfigure[US, rebalance every 15 days. Sharpe ratios of ACGAN, CGAN, and Markowitz are \textbf{0.82}, 0.49, and 0.56 respectively.]{\includegraphics[width=0.325\textwidth]{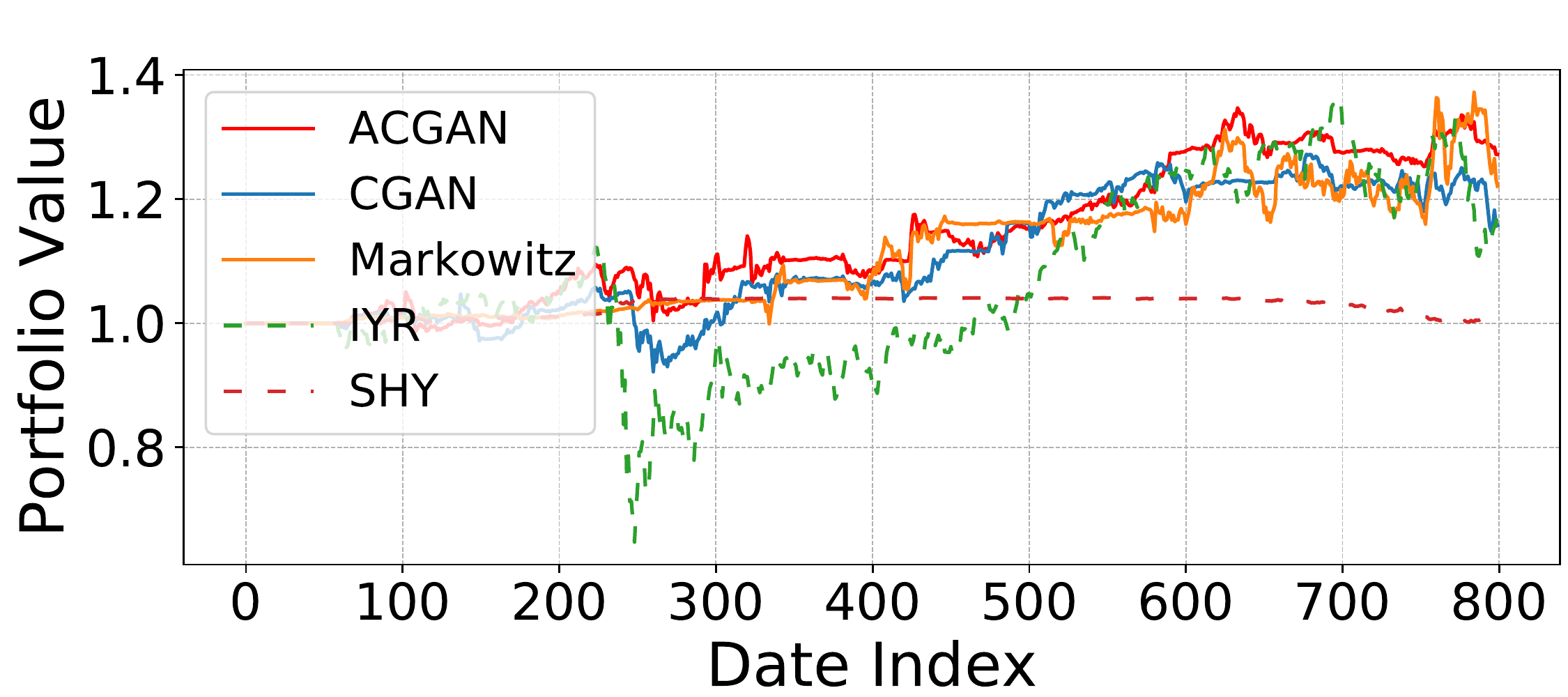} \label{fig:portfolio_value_us2}}
\subfigure[US, rebalance every 20 days.
Sharpe ratios of ACGAN, CGAN, and Markowitz are \textbf{1.20}, 0.71, and 0.45 respectively. ]{\includegraphics[width=0.325\textwidth]{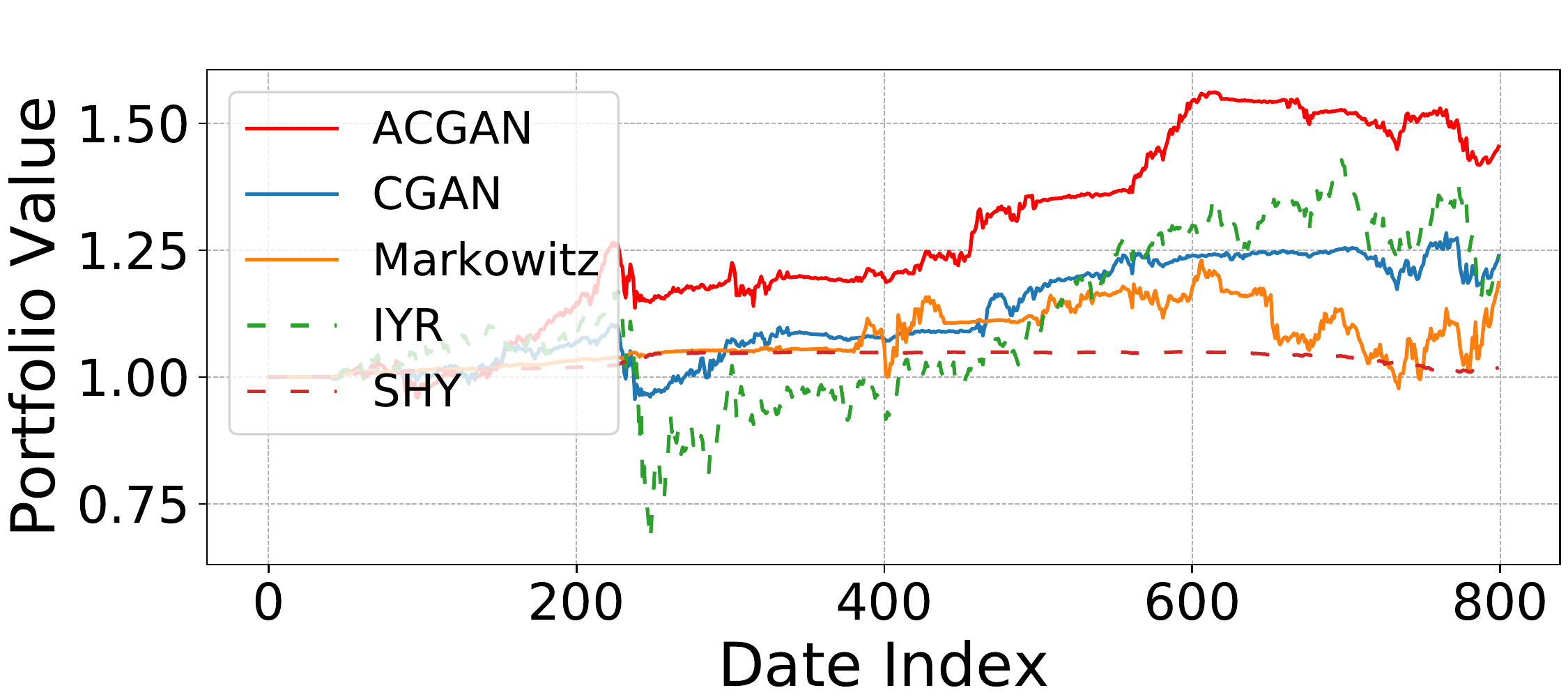} \label{fig:portfolio_value_us3}}
\subfigure[EU, rebalance every 10 days, 
Sharpe ratios of ACGAN, CGAN, and Markowitz are \textbf{1.02}, 0.95, and, 0.74 respectively.
]{\includegraphics[width=0.325\textwidth]{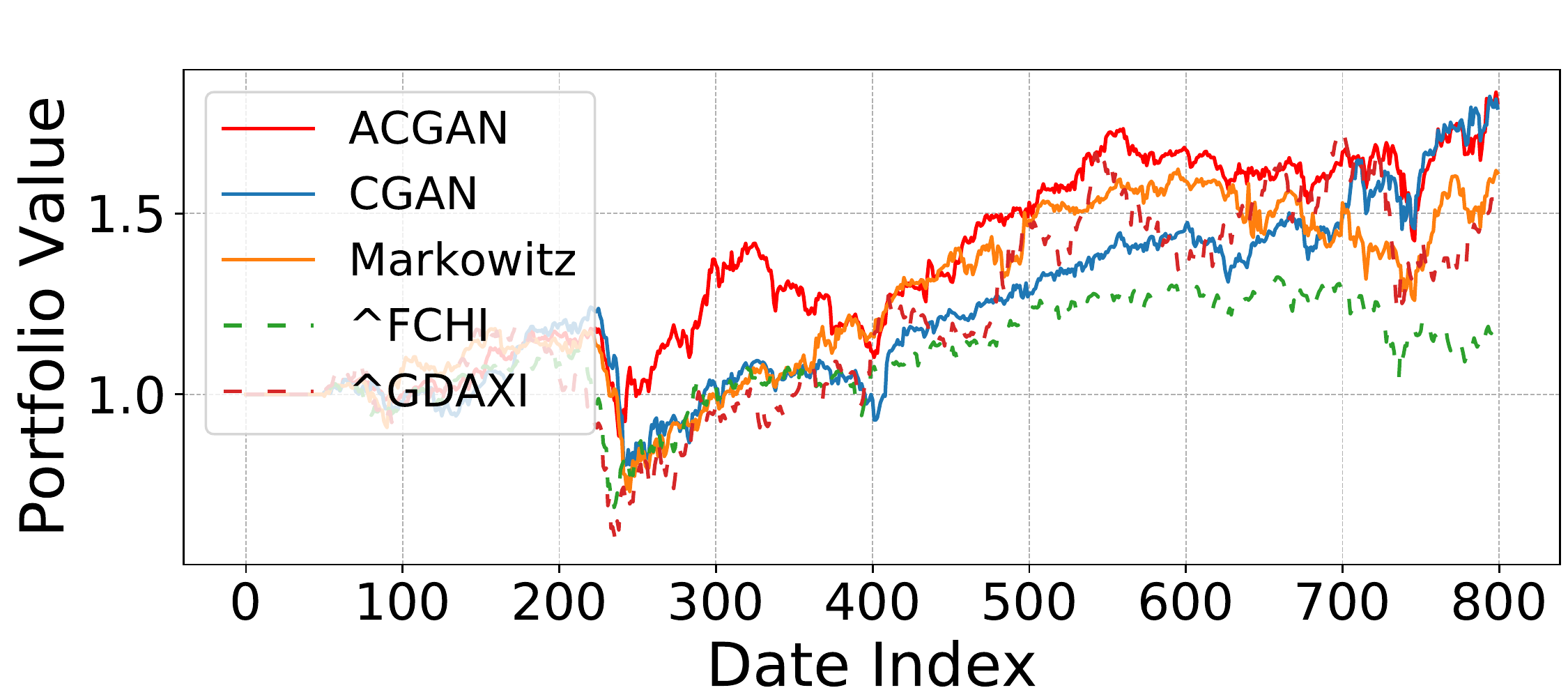} \label{fig:portfolio_value_eu1}}
\subfigure[EU, rebalance every 15 days. Sharpe ratios of ACGAN, CGAN, and Markowitz are \textbf{0.96}, 0.76, and 0.64 respectively.]{\includegraphics[width=0.325\textwidth]{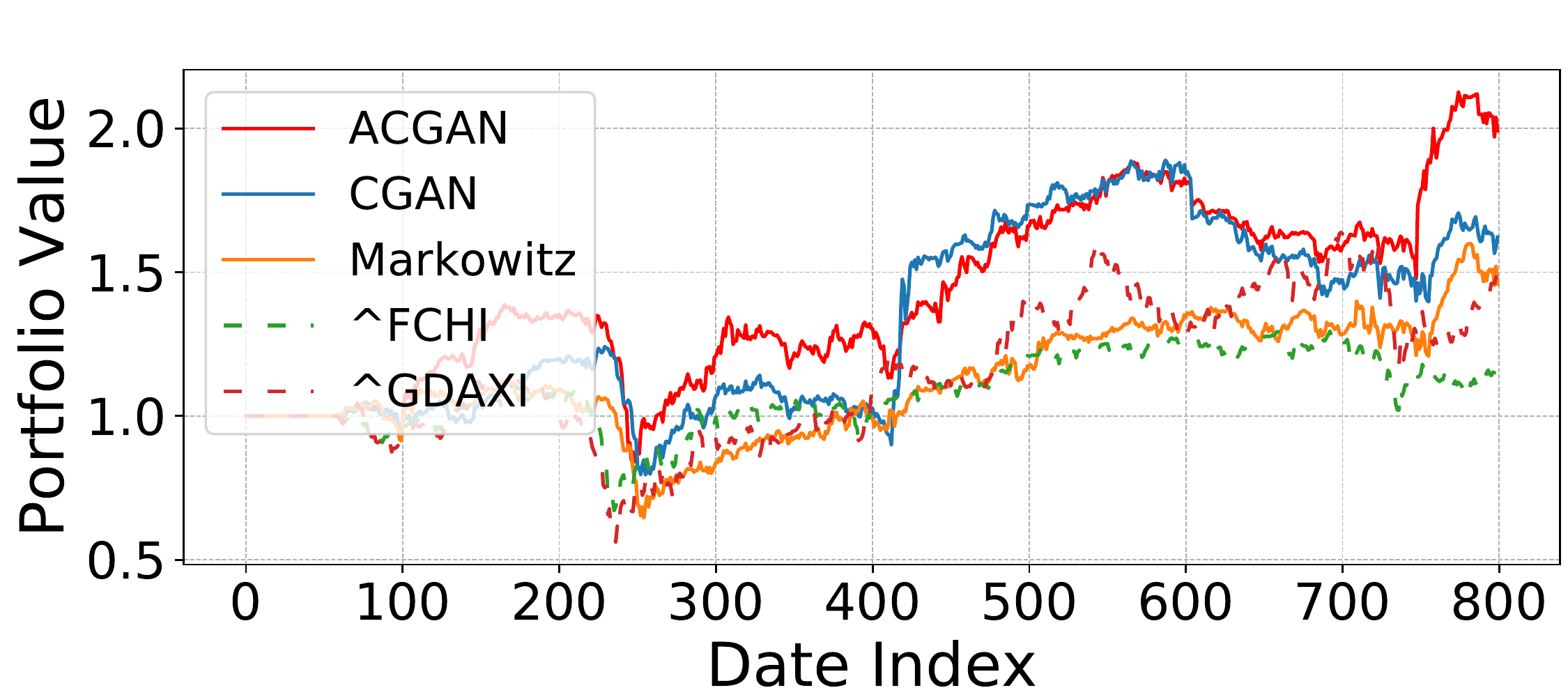} \label{fig:portfolio_value_eu2}}
\subfigure[EU, rebalance every 20 days. 
Sharpe ratios of ACGAN, CGAN, and Markowitz are \textbf{1.27}, 1.05, and 1.06 respectively.
]{\includegraphics[width=0.325\textwidth]{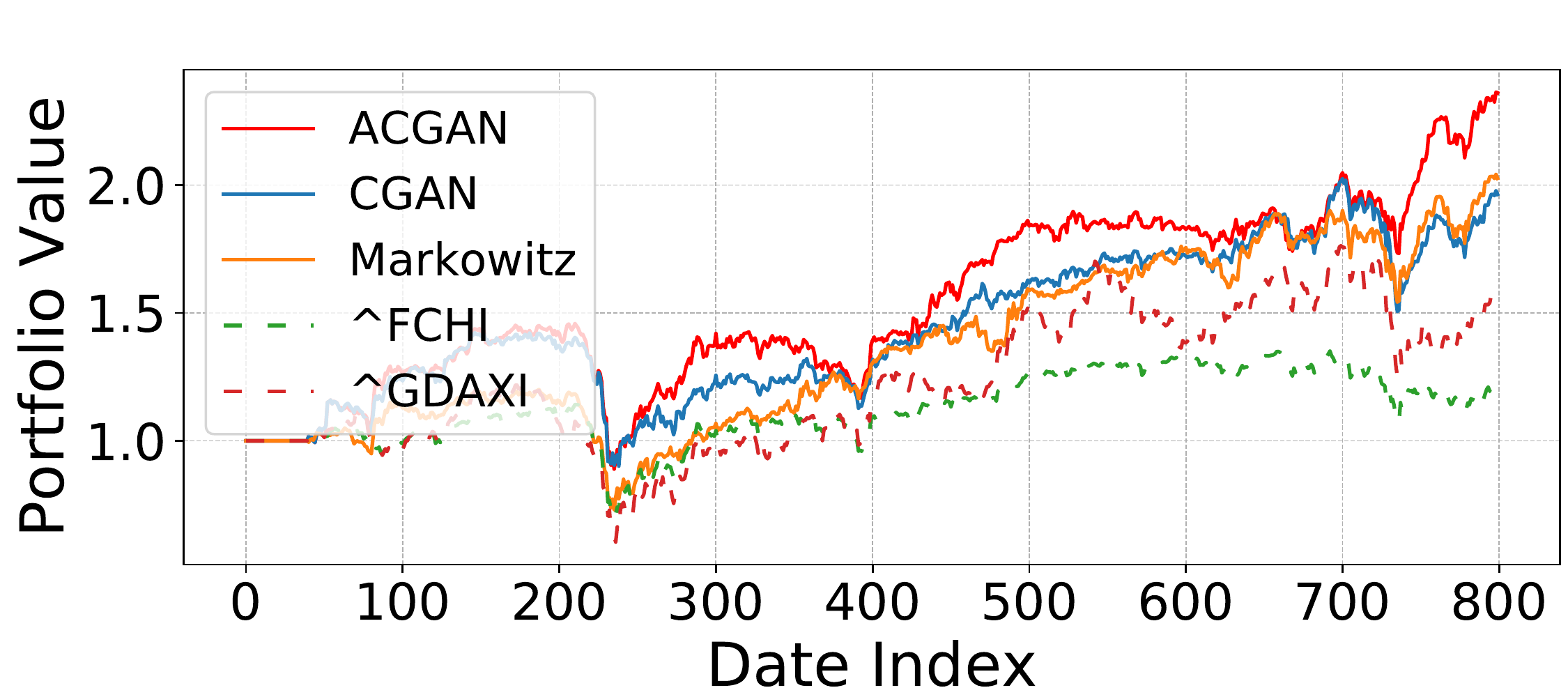} \label{fig:portfolio_value_eu3}}
\caption{Portfolio values for different diversification risk settings. Reference benchmarks are shown with dashed lines (Index or ETF assets). CGAN, ACGAN, and Markowitz with solid lines.}
\label{fig:portfolio_value_acgan_compare}
\end{figure*}

\subsection{Generating Analysis}
We follow the training and testing procedures in Algorithm~\ref{alg:acgan_training}. Given the training matrix $\bM$ of size $N\times D$ (where $N$ is the number of assets and $D$ is the number of days in the daily analysis context) and the window size $w$ ($w=h+f$ where $h$ is the length of the historical window and $f$ is the length of the future window), we define the index set $\mathcal{S}_1=\{1,2,\ldots, D-w+1\}$ so that $D-w$ samples can be extracted for each training epoch. While at the the testing stage, given the testing matrix $\bX\in \real^{N\times K}$, the index set is obtained by $\mathcal{S}_2=\{\textcolor{black}{h+1},h+f+1,\ldots\}$ so that $(K-h)/f$ samples can be obtained (supposed here $(K-h)$ can be divided by $f$). 
The output $\bY\in \real^{N\times K}$ of Algorithm~\ref{alg:acgan_training} is financial market simulation of the $N$ assets in $K$ days (here $N=10$ and $K=800$ in our datasets from US and EU regions).
To be more concrete, the first $h$ days of $\bY$ are just copy of $\bX$, while the next $f$ days are the synthetic series based on the data of the first $h$ days; the next $f$ days are the synthetic series based on the data between the $f$-th and $(f+h)$-th days; and so on.

We set window size $h=40, f=20$ and $w=60$ in all experiments.
Figure~\ref{fig:diversity_us_eu_ecgan} shows the actual price trend (black solid line) of the US assets and the EU assets for the first 100 trading days in the test set, and five representative simulations generated by ACGAN (colored dashed lines). The simulated price series of the whole period (800 trading days) and simulated price series of the CGAN model (100 and all trading days) can be found in Figure~\ref{fig:diversity_us_eu_acgan_800}, \ref{fig:diversity_us_eu_cgan}, and \ref{fig:diversity_us_eu_cgan_800} respectively.

We then calculate the Pearson correlation between the real series and the synthetic series by ACGAN and CGAN models. The Pearson correlation between two series $\bx$ and $\by$ is defined by
$
\text{Corr}(\bx,\by) =\frac{\mathbb{E}[(\bx-\mu_x)(\bx-\mu_y)]}{\sigma_x\sigma_y},
$
where $\mu_x, \sigma_x$ are the mean and standard deviation of $\bx$.
We generate 1,000 synthetic series. Table~\ref{table:correlation_acgan-cgan} presents the mean correlation for different assets. In most cases, the correlation between the real series and synthetic series by ACGAN achieves larger values making the ACGAN a better model to generate financial series closer to the real ones. While there exist other correlation measurements, e.g., the Cosine correlation, and time-varying correlation \citep{tulchinsky2019finding}, the results do not alter significantly.

Moreover, we analyze different statistical properties of the generated series as discussed in Section~\ref{section:statistica_property_financial_tim} \footnote{Other than these single-asset properties, we also report some cross-asset statistical properties in Appendix~\ref{section:cross_statistics_ACGAN}.}. 
Figure~\ref{fig:autocorrelation_us_fake_0} shows the decay of the autocorrelation function of the price return on a daily timeframe (for GOOG asset in the US market). The absence of linear correlation in the price return on a daily timeframe suggests that the generated financial series are efficient to a certain extent. Similarly, the fat-tail, leverage effect, and coarse-fine properties of the generated data by CGAN and ACGAN all satisfy the statistical properties we have mentioned previously. Table~\ref{table:statistics_property-cgan-acgan} summarizes the coefficients of the real market data and data generated by CGAN and ACGAN, where we also include the Kurtosis and skewness of these assets. In practice, the Kurtosis of a normal distribution is 3 and the value is larger than 3 for a fat-tailed distribution. In most cases, the coefficients of the ACGAN are closer to the true ones.

\subsection{Portfolio Analysis}

After generating the synthetic series for each asset, we optimize over the fake series to generate minimal Sharpe ratio weight allocations (for ACGAN and CGAN). For Markowitz framework, the optimization is done over the past data.  
We consider three rebalance settings: a \textit{defensive setting} with rebalancing every $\eta=10$ days; a \textit{balanced setting} with $\eta=15$;
and an \textit{aggressive setting} with $\eta=20$.
Figure~\ref{fig:risk-sharperation_acgan_compare} presents the distribution of return-SR (Sharpe ratio) scatters with 1,000 draws from ACGAN and CGAN models, and the one from Markowitz framework. 
The \textit{ACGAN (Mean)} and \textit{CGAN (Mean)} are strategies by taking the average weight from these 1,000 draws on each rebalancing date.
The points in the upper-right corner are the better ones. 
In all scenarios, the mean strategies of the proposed ACGAN model have better returns and Sharpe ratios than those of the Markowitz model and the CGAN model. 
When we apply the mean strategy in the US region, the ACGAN achieves both better return and Sharpe ratio evaluations compared to the mean strategy of CGAN.
Similar results are observed in the EU region with balanced and aggressive settings with $\eta=15,20$. 
The return results of the ACGAN and CGAN models are close in the defensive setting for the EU region; while ACGAN has better Sharpe ratio such that it takes smaller risk.
 

Figure~\ref{fig:portfolio_value_acgan_compare} shows the portfolio value series of mean strategies for ACGAN and CGAN, and the one from Markowitz framework along the test period where we initialize each portfolio with a unitary value. ACGAN dominates the other approaches in terms of the final portfolio value.

\section{Conclusion}
The aim of this paper is to solve the issue of poor prediction ability in the CGAN methodology for portfolio analysis. We propose a simple and computationally efficient algorithm that requires little extra computation and is easy to implement for conditional time series generation. Overall, we show that the proposed ACGAN model is a versatile framework that synthesizes series with larger correlation to the true time series and thus yields better portfolio allocation. ACGAN is able to keep as much original information as possible while still enables the generator to generate data closer to the ones receiving high scores from discriminator. 
While it still remains interesting if the ACGAN model can be applied in computer vision context to generate images that are closer to the real ones.

\paragraph{Acknowledgments}
We greatly appreciate insightful discussions with Giovanni Mariani on the data normalization and the framework of the PAGAN (CGAN) methodology. 


\bibliography{bib}
\bibliographystyle{sty}
\balance

\onecolumn
\appendix

\section{Network Structures}\label{appendix:acgan_net_structures}
We provide detailed structure for the neural network architectures we used in our experiments in this section.
Given the number of assets $N$, historical length $h$, future length $f$ ($w=h+f$), and latent dimension $m$ for the prior distribution vector $\bz$, we consider multi-layer perceptron (MLP) structures, 
the detailed architecture 
for each fully connected
layer is described by F$(\langle \textit{num inputs} \rangle :\langle \textit{num outputs} \rangle:\langle \textit{activation function} \rangle)$; 
for an activation function of LeakyRelu with parameter $p$ is described by $\text{LR}(\langle p \rangle )$; 
and for a dropout layer is described by
DP$(\langle \textit{rate} \rangle)$. The \textit{conditioner} in CGAN shares the same structure as the \textit{encoder} in the ACGAN model (see Figure~\ref{fig:cgan_structure} and Figure~\ref{fig:acgan_structure}).
Then the network structures we use can be described as follows:

\begin{align}
&\textbf{Conditioner}=\textbf{Encoder}=
\text{F\big($N\cdot h$: 512: LR(0.2)\big)} \cdot 
\text{F\big(512:512:LR(0.2)\big)} \cdot 
\text{DP(0.4)}\cdot
\text{F(512:16)}\\
&\textbf{Decoder}=
\text{F\big(16:512:LR(0.2)\big)} \cdot 
\text{F\big(512:512:LR(0.2)\big)} \cdot 
\text{DP(0.4)}\cdot
\text{F(512:$N\cdot h$)}\\
&\textbf{Simulator}=
\text{F\big($m$+16:128:LR(0.2)\big)} \cdot 
\text{F\big(128:256:LR(0.2)\big)} \cdot \nonumber\\
&\gap\gap\gap\gap\gap\gap\gap\gap \text{  F\big(256:512:LR(0.2)\big)} \cdot
\text{F\big(512:1024:LR(0.2)\big)} \cdot
\text{F\big(1024:$N\cdot f$:TanH\big)} \\
&\textbf{Discriminator}=
\text{F\big($N\cdot (h$+$f)$:512:LR(0.2)\big)} \cdot 
\text{F\big(512:512:LR(0.2)\big)} \cdot \nonumber\\
&\gap\gap\gap\gap\gap\gap\gap\gap \text{  DP(0.4)} \cdot 
\text{F\big(256:512:LR(0.2)\big)} \cdot
\text{F(512:1)}
\end{align}

We trained networks using Adam's optimizer with learning rate $2\times 10^{-5}$, $\beta_1 = 0.5$, and $\beta_2=0.999$. We set the penalization parameters $\lambda_1=10$ and $\lambda_2=3$. The latent dimension is $m=100$.
And we trained models for 1,000 epochs. 
%



\section{More Samples for the Generative Models}
\begin{figure}[H]
\centering  
\subfigtopskip=2pt 
\subfigbottomskip=2pt 
\subfigcapskip=-2pt 
\subfigure{\includegraphics[width=1\textwidth]{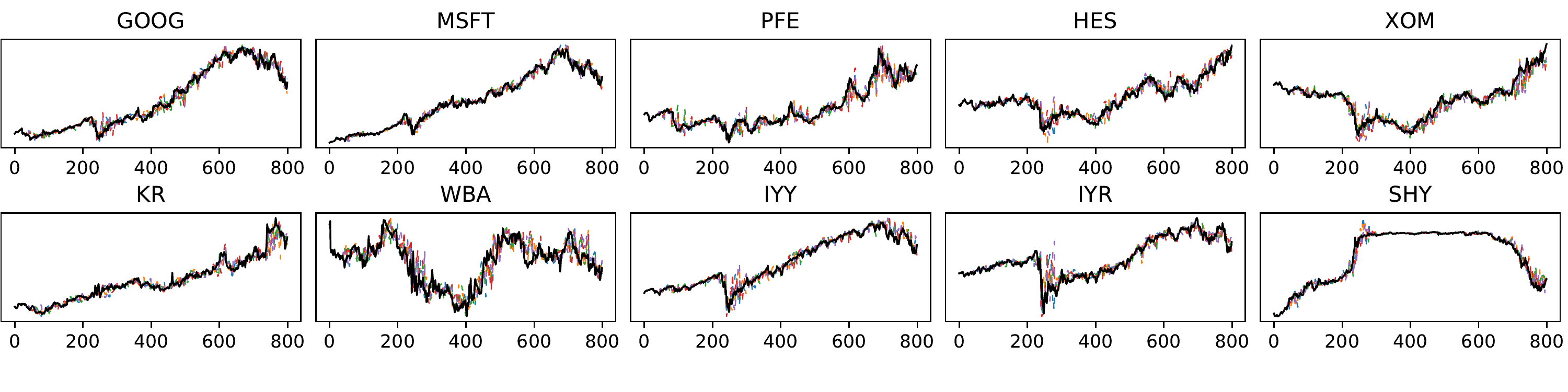}}
\subfigure{\includegraphics[width=1\textwidth]{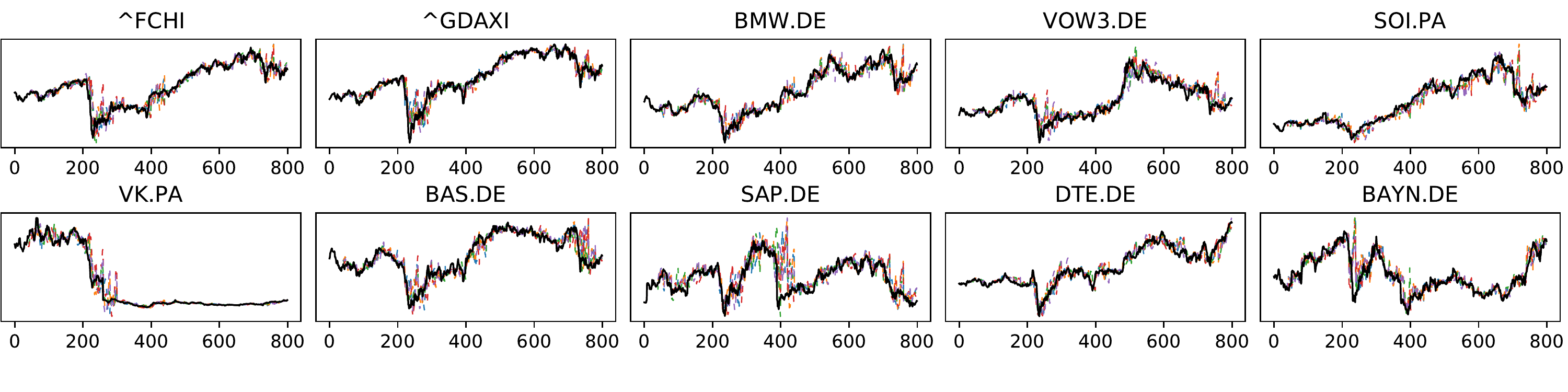}}
\caption{\textbf{ACGAN:} Actual price trend (black solid line) of the US assets (upper) and the EU assets (lower) for 800 trading days in the test set (the whole period of the test set), and five representative simulations generated by ACGAN (colored dashed lines).}
\label{fig:diversity_us_eu_acgan_800}
\end{figure}

\begin{figure}[H]
	\centering  
	\subfigtopskip=2pt 
	\subfigbottomskip=2pt 
	\subfigcapskip=-2pt 
	\subfigure{\includegraphics[width=1\textwidth]{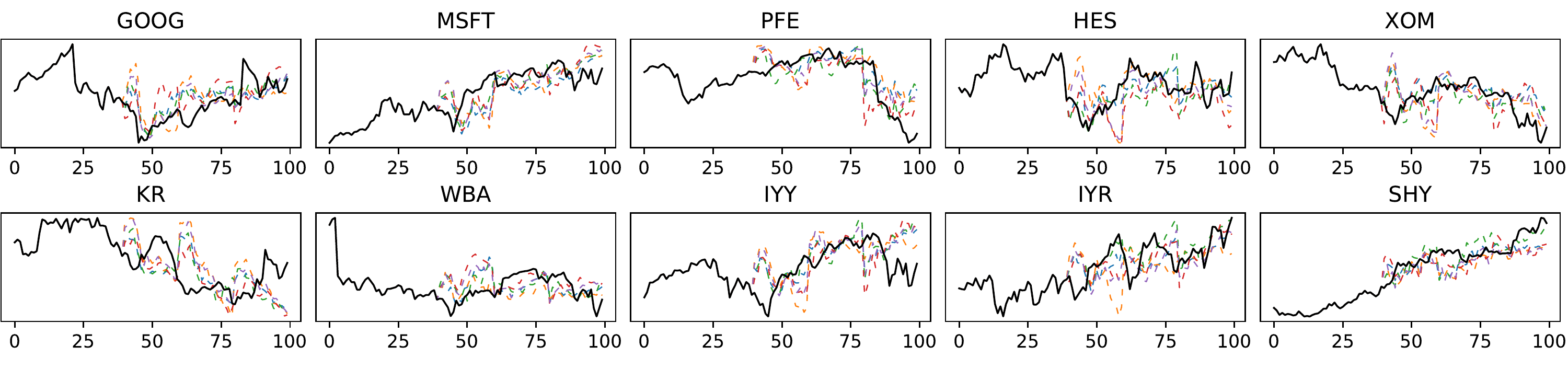}}
	\subfigure{\includegraphics[width=1\textwidth]{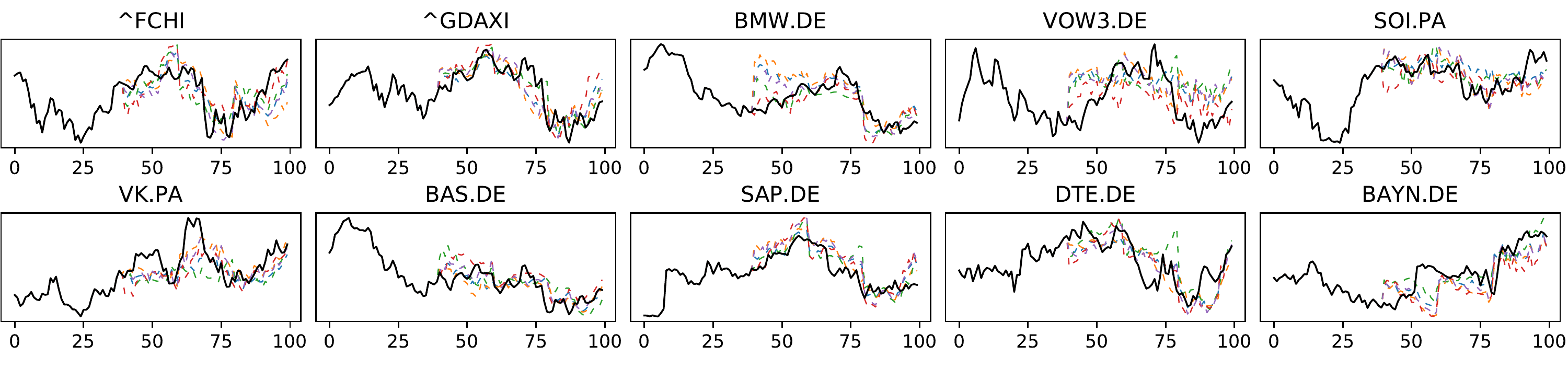}}
	\caption{\textbf{CGAN:} Actual price trend (black solid line) of the US assets (upper) and the EU assets (lower)  for the first 100 trading days in the test set, and five representative simulations generated by CGAN (colored dashed lines).}
	\label{fig:diversity_us_eu_cgan}
\end{figure}

\begin{figure}[H]
	\centering  
	\subfigtopskip=2pt 
	\subfigbottomskip=2pt 
	\subfigcapskip=-2pt 
	\subfigure{\includegraphics[width=1\textwidth]{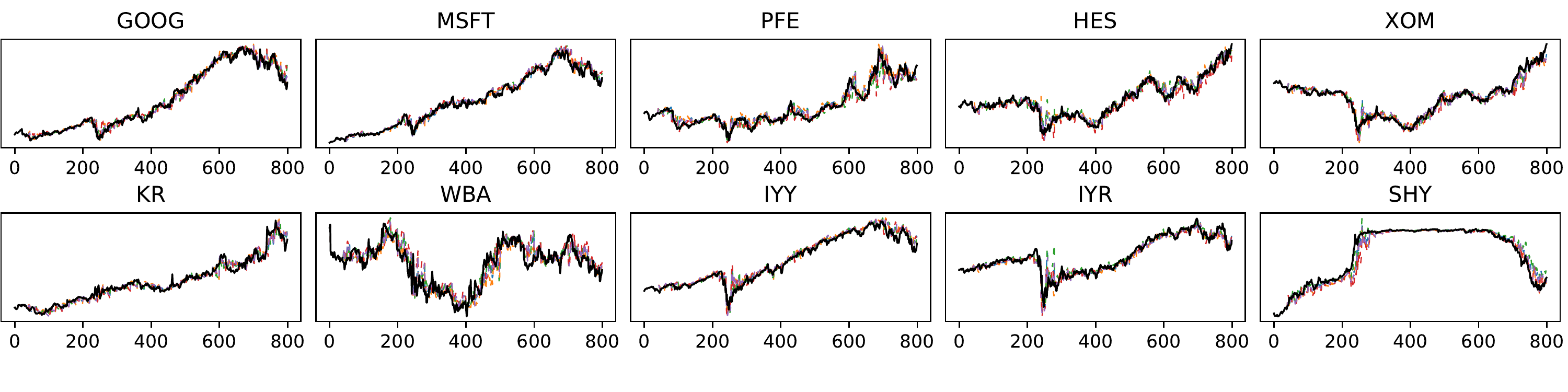}}
	\subfigure{\includegraphics[width=1\textwidth]{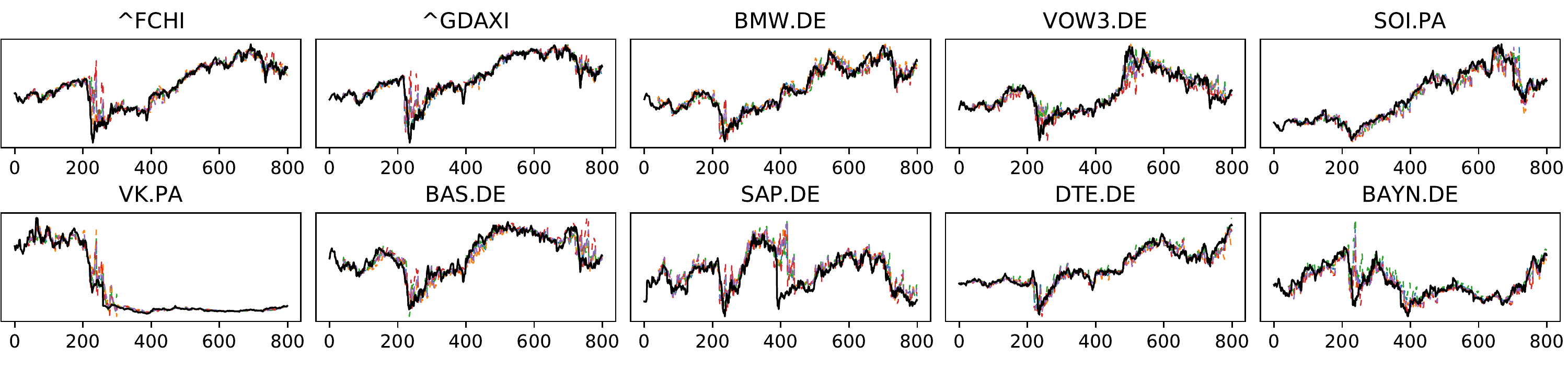}}
	\caption{\textbf{CGAN:} Actual price trend (black solid line) of the US assets (upper) and the EU assets (lower) for 800 trading days in the test set (the whole period of the test set), and five representative simulations generated by CGAN (colored dashed lines).}
	\label{fig:diversity_us_eu_cgan_800}
\end{figure}

In this section, we provide move synthetic results for both ACGAN and CGAN models. 
We set window size $h=40, f=20$ and $w=60$ in all experiments.
As shown in the main paper, Figure~\ref{fig:diversity_us_eu_ecgan} presents the actual price trend (black solid line) of the US assets and the EU assets for the first 100 trading days in the test set, and five representative simulations generated by ACGAN (colored dashed lines).
Figure~\ref{fig:diversity_us_eu_acgan_800} reports the simulated price series of the whole period (800 trading days). 
And simulated price series of the CGAN model (100 and all trading days) can be found in \ref{fig:diversity_us_eu_cgan}, and \ref{fig:diversity_us_eu_cgan_800} respectively.

\section{Statistical Properties for Multiple Assets}\label{section:cross_statistics_ACGAN}

Following \citet{cont2001empirical}, we also consider the stylized facts across different assets, namely the cross-asset correlation, volatility correlation, and cross-asset leverage effect. 

\paragraph{Cross-Asset Correlation}
Given two assets $i,j$ whose return values at time $t$ are denoted by $r_{i,t}, r_{j,t}$ respectively. \textit{Cross-asset correlation} considers the correlation of two assets at different time frames. Under weak market efficiency, there should be no obvious lead-lag cross correlation between different assets; otherwise, it will provide arbitrage space:
$$
\text{Corr}(r_{\textcolor{blue}{i},t}, r_{\textcolor{blue}{j},t+k}) =
\frac{\mathbb{E} [(r_{\textcolor{blue}{i}, t}-\mu_i)(r_{\textcolor{blue}{j}, t+k}-\mu)] }{\sigma_i\sigma_j} \approx 0, \forall\,\, k\geq 2,
$$
where $\mu_i, \sigma_i$ are the mean and standard deviation of return series $\br_i$ of asset $i$.
In practice, the cross correlation presents positive correlation for the 0-th and 1-th lag values; and no significant correlation for higher order values. Table~\ref{table:statistics_property-cgan-acgan_crossproperty} considers the mean coefficients of the first 10 orders across different combinations of assets. For 10 assets, there are 45 such combinations.

\paragraph{Volatility Correlation}
The \textit{volatility correlation} consider the lead-lag cross correlation between the absolute values of different asset returns:
$$
\text{Corr}(|r_{\textcolor{blue}{i},t}|, |r_{\textcolor{blue}{j},t}|).
$$
Although there is no significant cross correlation between different asset return series under the efficient markets hypothesis, its volatility may show a significant cross correlation. 
We may observe that the increasing in the volatility of one asset can lead to the rise of the volatility of another asset. In the short term, there is a large positive correlation between volatilities of different assets; and the cross correlation will gradually decline over time.
Table~\ref{table:statistics_property-cgan-acgan_crossproperty} considers the mean coefficients of the first 10 orders across different combinations of assets.

\paragraph{Cross-Asset Leverage Effect}
Similar to the single-asset leverage effect (Section~\ref{section:statistica_property_financial_tim}), the \textit{cross-asset leverage effect} means there is a negative correlation between the past price returns of one asset and the future volatility of another asset:
$$
L_{ij}(k) = \frac{\mathbb{E}[r_{\textcolor{blue}{i},t} |r_{\textcolor{blue}{j},t+k}|^2] - 
\mathbb{E}[	r_{\textcolor{blue}{i},t}]\cdot  \mathbb{E}[|r_{\textcolor{blue}{j},t}|^2]}{\mathbb{E}[|r_{\textcolor{blue}{j},t}|^2]^2}.
$$
Again, Table~\ref{table:statistics_property-cgan-acgan_crossproperty} considers the mean coefficients of the first 10 orders across different combinations of assets.

We can observe from Table~\ref{table:statistics_property-cgan-acgan_crossproperty} that the cross-asset correlation and volatility correlation of ACGAN and CGAN are close. However, ACGAN  shows promise in terms of the cross-asset leverage effect.

\begin{table*}[]
\begin{tabular}{l|lll|lll}
\hline
Statistics      & \gap Real (US) & ACGAN (US) & CGAN (US) & \gap Real (EU) & ACGAN (EU) & CGAN (EU) \\ \hline
Cross correlation & \gap  0.0204 &  \textbf{ 0.0223 }  &  \textbf{ 0.0219 } & \gap 0.0190  & \textbf{ 0.0205  } & \textbf{  0.0191 }\\
Volatility correlation& \gap  0.1855  & \textbf{ 0.1872 }  &  0.1748   &\gap  0.1279   & \textbf{ 0.1410  }  & \textbf{ 0.1405 }   \\
Cross leverage effect  & \gap  -10.2414   & \textbf{ -6.5860 }   &  0.9129  &\gap  -6.4018  & \textbf{ -3.6489  } &   -2.9875  \\
\hline
\end{tabular}
\caption{Cross statistical properties of the real and generated series. Cross correlation value is the mean of the first 10-th lags of cross correlation coefficients (across different combinations of assets); Similarly, the volatility correlation and cross leverage effect shown in the table consider the mean of first 10-th lags of leverage effect coefficients (across different combinations of assets). ACGAN shows promise in terms of the cross leverage effect.}
\label{table:statistics_property-cgan-acgan_crossproperty}
\end{table*}

\end{document}